\documentclass[twocolumn,showpacs,aps,epsfig]{revtex4}

\usepackage{graphicx}
\usepackage{epstopdf}
\usepackage{latexsym}

\usepackage[center]{subfigure}

\begin{document}

 \newcommand{\bq}{\begin{equation}}
 \newcommand{\eq}{\end{equation}}
 \newcommand{\bqn}{\begin{eqnarray}}
 \newcommand{\eqn}{\end{eqnarray}}
 \newcommand{\nb}{\nonumber}
 \newcommand{\lb}{\label}
\newcommand{\PRL}{Phys. Rev. Lett.}
\newcommand{\PL}{Phys. Lett.}
\newcommand{\PR}{Phys. Rev.}
\newcommand{\CQG}{Class. Quantum Grav.}

\title{Brane cosmology  in the   Horava-Witten heterotic M-Theory on $S^{1}/Z_{2}$}

\author{Qiang Wu $^{1, 2}$}
\email{qiang_wu@baylor.edu}
\author{ Yungui Gong ${}^{3}$}
\email{gongyg@cqupt.edu.cn}
\author{Anzhong Wang ${}^{2}$}
\email{anzhong_wang@baylor.edu}
\affiliation{ $^{1}$  Department of Physics, Zhejiang University of
Technology,
Hangzhou 310032,  China\\
${}^{2}$ GCAP-CASPER, Physics Department, Baylor University,
Waco, TX 76798-7316\\
${}^{3}$ College of Mathematics $\&$ Physics, Chongqing University
of Posts $\&$
Telecommunications, Chongqing 400065,   China}

\date{\today}
\begin{abstract}

We study the radion stability and radion mass in the framework 
of the Horava-Witten (HW) heterotic M-Theory on $S^{1}/Z_{2}$, 
and find that the radion is stable and its mass can be of the 
order of GeV. The gravity is localized on the visible 
brane, and the spectrum of the gravitational Kaluza-Klein (KK) 
modes is discrete and can have a mass gap of TeV. The 
corrections to the 4D Newtonian potential from the higher order 
gravitational KK modes are exponentially suppressed. Applying 
such a setup to cosmology, we find the generalized Friedmann-like 
equations on each of the two orbifold branes.

\end{abstract}

\maketitle

\section{Introduction}

\renewcommand{\theequation}{1.\arabic{equation}} \setcounter{equation}{0}

Recent observations of supernova (SN) Ia reveal the striking discovery that
our universe has lately been in its accelerated expansion phase \cite{agr98}. 
Cross checks from the cosmic microwave background radiation and large
scale structure all confirm this unexpected result \cite{Obs}. Such an
expansion was predicted neither by the standard model of particle physics
nor by the standard model of cosmology.

In Einstein's theory of gravity, in order to have such an acceleration, it
demands the introduction of either a tiny positive cosmological constant or
an exotic component of matter, which has a very large negative pressure and
interacts with other components of matter weakly. This
invisible component is usually dubbed as dark energy.

A tiny cosmological constant is well consistent with all observations
carried out so far \cite{SCH07}, and could represent one of the simplest
resolutions of the crisis. However, considerations of its origin lead to
other severe problems: (a) Its theoretical expectation values exceed
observational limits by $120$ orders of magnitude \cite{wen}. Even if such
high energies are suppressed by supersymmetry, the electroweak corrections
are still $56$ orders higher. (b) Its corresponding energy density $%
\rho_{\Lambda} \equiv \Lambda/(8\pi G)$ is comparable with that of matter
only recently. Otherwise, galaxies would have not been formed. Considering
the fact that the energy density of matter depends on time, one has to
explain why only \emph{now} the two are in the same order. (c) Once the
cosmological constant dominates the evolution of the universe, it dominates
forever. An eternally accelerating universe seems not consistent with
string/M-Theory, because it is endowed with a cosmological event horizon
that prevents the construction of a conventional S-matrix describing
particle interaction \cite{Fish}. Other problems with an asymptotical de
Sitter universe in the future were further explored in \cite{KS00}.

In view of all the above, dramatically different models have been proposed,
including quintessence \cite{quit}, the DGP branes \cite{DGP00}, and $f(R)$
models \cite{FR}. For details, see \cite{DE} and references therein. However,
it is fair to say that so far no convincing model has been constructed.

Since the cosmological constant problem is intimately related to quantum gravity, its
solution is expected to come from quantum gravity, too. At the present,
string/M-Theory is our best bet for a consistent quantum theory of gravity,
so it is reasonable to ask what string/M-Theory has to say about the
cosmological constant. In the string
landscape \cite{Susk}, it is expected there are many different vacua with
different local cosmological constants \cite{BP00}. Using the anthropic
principle, one may select the low energy vacuum in which we can exist.
However, many theorists still hope to explain the problem without invoking
the existence of ourselves. In addition, to have a late time accelerating
universe from string/M-Theory, Townsend and Wohlfarth \cite{townsend}
invoked a time-dependent compactification of pure gravity in higher
dimensions with hyperbolic internal space to circumvent Gibbons' non-go
theorem \cite{gibbons}. Their exact solution exhibits a short period of
acceleration. The solution is the zero-flux limit of spacelike branes 
\cite {ohta}. If non-zero flux or forms are turned on, a transient acceleration
exists for both compact internal hyperbolic and flat spaces \cite{wohlfarth}%
. Other accelerating solutions by compactifying more complicated
time-dependent internal spaces can be found in \cite{string}.

Recently, we \cite{GWW07} studied the cosmological constant problem and the
late transient acceleration of the universe in the framework of the
Horava-Witten heterotic M-Theory on $S^{1}/Z_{2}$ \cite{HW96}. Using the
Arkani-Hamed-Dimopoulos-Dvali (ADD) mechanism of large extra dimensions \cite
{ADD98}, it was shown that the effective cosmological constant on each of
the two branes can be  lowered to its current observational value. The
domination of this term is only temporary. Due to the interaction of the
bulk and the brane, the universe will be in its decelerating expansion phase
again, whereby all problems connected with a far future de Sitter universe
\cite{Fish,KS00} are resolved. 

Such studies were further generalized to string
theory \cite{WS07,WSVW08,WS09}, and were showed that the same mechanism is also
viable in all of the five versions of string theory. In addition, the radion
stability was also investigated, by using the Goldberger-Wise mechanism \cite
{GW99}, and showed explicitly that it is stable. 

In this paper, we shall give a systematical study of brane worlds in the
framework of the Horava-Witten (HW) heterotic M-Theory on $S^{1}/Z_{2}$
\cite{HW96,LOSW99}. We first address two important issues, which are fundamental   
in order for the model to be viable: (i) the radion stability and its mass; 
and (ii) the localization 
of gravity, the 4D effective Newtonian potential and its corrections from the high
order gravitational KK modes.  Then, 
we apply the
model to cosmology, and write down explicitly the general gravitational and matter 
field equations both in the bulk and on the branes. In particular, the paper is 
organized as follows: In Sec. II, we consider the  HW heterotic M-Theory on 
$S^{1}/Z_{2}$ along the line set up by Lukas {\em et al.} in \cite{LOSW99}.
To consider its cosmological applications, we add a potential term and matter
fields on each of the two branes. 
In Sec. III, we consider the
radion stability and radion mass, using the Goldberger-Wise mechanism \cite{GW99}.
In Sec. IV, we study the localization of gravity and calculate the 4-dimensional 
effective Newtonian potential. The spectrum of gravitational Kaluza-Klein (KK) modes is 
worked out explicitly, and found to be discrete and can have a mass gap of TeV.
In Sec. IV, applying the model to cosmology, we separate the gravitational and matter field
equations into two group, one holds outside of the two branes, and one holds on 
each of the two branes. In particular, we find the most general generalized Friedmann-like 
equations on each of the two orbifold branes. The paper is ended with Sec. V, in which
our main findings are summarized, and some discussing remarks are given.

It should be noted that brane worlds have  been studied intensively in the past decade 
\cite{branes}. However, to our best knowledge, such studies in the HW setup 
have not been carried out in details \cite{Chen06}. 

It is also interesting to note that in 4-dimensional spacetimes there exists 
Weinberg's no-go theorem for the adjustment of the cosmological constant \cite{wen}. 
However, in higher dimensional spacetimes, the 4-dimensional vacuum energy on 
the brane does not necessarily give rise to an effective 4-dimensional cosmological
constant. Instead, it may only curve the bulk, while leaving the brane still
flat \cite{CEG01}, whereby Weinberg's no-go theorem is evaded. It was
exactly in this vein, the cosmological constant problem was studied in the
framework of brane worlds in 5-dimensional spacetimes \cite{5CC} and
6-dimensional supergravity \cite{6CC}. However, it was soon found that in
the 5-dimensional case hidden fine-tunings are required \cite{For00}. In the
6-dimensional case such fine-tunings may not be needed, but it is still not
clear whether loop corrections can be as small as expected \cite{Koyama07,Burg07}.

\section{The Model}

\renewcommand{\theequation}{2.\arabic{equation}} \setcounter{equation}{0}

Let us consider the 11-dimensional spacetime of the Horava-Witten M-Theory,
described by the metric \cite{LOSW99},
\begin{equation}  \label{2.1}
ds^{2}_{11} = V^{-2/3} g_{ab} dx^{a}dx^{b} - V^{1/3}\Omega_{ij}dz^{i}dz^{j},
\end{equation}
where $ds^{2}_{CY,6} \equiv \Omega_{ij}dz^{i}dz^{j}$ denotes the Calabi-Yau
3-fold, and $V $ is the Calabi-Yau volume modulus that measures the
deformation of the Calabi-Yau space, and depends only on $x^{a}$, where $a =
0, 1, ... , 4$.

Note that in this paper we shall use some notations slightly different from
the ones used in \cite{GWW07}.

\subsection{5-Dimensional Effective Actions}

By integrating the corresponding 11-dimensional action over Calabi-Yau
3-fold, the 5-dimensional effective action of the Horava-Witten theory is
given by \cite{LOSW99}
\begin{eqnarray}  \label{2.2}
S_{5} &=& - \frac{1}{2\kappa^{2}_{5}}\int_{M_{5}}{\sqrt{g} \left(R[g]
- \frac{1}{2} \left(\nabla\phi\right)^{2} + 6\alpha^{2} e^{-2\phi}\right)}
\nonumber\\
& & - \sum_{I= 1}^{2}{\epsilon_{I} \frac{6\alpha}{\kappa^{2}_{5}}%
\int_{M^{(I)}_{4}} {\sqrt{-g^{(I)}} e^{-\phi}}},
\end{eqnarray}
where $I = 1, \; 2,\; \epsilon_{1} = - \epsilon_{2} = 1$, $\nabla$ denotes
the covariant derivative with respect to $g_{ab}$, and
\begin{equation}  \label{2.2a}
\phi \equiv \ln({V}),\;\;\; \kappa^{2}_{5} \equiv \frac{\kappa^{2}_{11}}{%
v_{CY,6}},
\end{equation}
with $v_{CY,6}$ being the volume of the Calabi-Yau space,
\begin{equation}  \label{2.2b}
v_{CY,6} \equiv \int_{X}{\sqrt{\Omega}}.
\end{equation}
The constant $\alpha$ is related to the internal four-form that has to be
included in the dimensional reduction \cite{LOSW99}. This four-form results from the
source terms in the 11-dimensional Bianchi identity, which are usually
non-zero. $g^{(I)}$ 's are the reduced metrics on the two boundaries $%
M^{(I)}_{4}$.

It should be noted that in general the dimensional reduction of the graviton
and the four-form flux generates a large number of fields \cite{LOSW99}. However, it is
consistent to set all the fields zero except for the 5-dimensional graviton
and the volume modulus. This setup implies that all components of the
four-form now point in the Calabi-Yau directions \cite{Chen06}. In addition, it can be
shown that the above action is indeed the bosonic sector of a minimal ${%
\mathcal{N}} =1$ gauged supergravity theory in 5-dimensional spacetimes
coupled to chiral boundary theories \cite{Ben99}.

To study cosmology in the above setup, we add matter fields on each of the
two branes,
\bqn
S_{4,\;m}^{(I)}&=& \int_{M_4^{(I)}}{\sqrt{-g^{(I)}}\left[\left({\mathcal{L}}%
_{4,m}^{(I)}\left(\phi ,\chi \right) -  V^{(I)}_{4}(\phi)\right)\right. }\nb\\
& & \;\;\;\;\; \left. - g^{(I)}_{k} \right],  
\label{2.3}
\eqn
where $\chi$ collectively denotes the SM fields localized on the branes,
$V^{(I)}_{4}(\phi)$ and $g^{(I)}_{k}$ are, respectively, the potential of
the scalar field  and the tension   of  the  I-th brane. As to be shown
below, $g^{(I)}_{k}$ is directly related to the four-dimensional Newtonian
constant $G_{4}$ \cite{Cline99}. 
 Clearly, these actions  in general make the 
two branes no longer supersymmetric, although the bulk still is. 

It should be noted that in general one also needs to include the Gibbons-Hawking
boundary term \cite{GH77} in the action (\ref{2.3}) \cite{Davis03}. However, 
here we work with the so-called upstairs  picture of the $S^{1}/Z_{2}$ orbifold 
\cite{KOS06}, where all total derivatives integrate to zero, while the boundary 
conditions are obtained by imposing the Lanczos equations \cite{Lan22}, as 
was done earlier by Israel \cite{Israel66}.

Variation of the action,
\begin{equation}
S_5^{total}=S_5+\sum_{I=1}^2{S_{4,\;m}^{(I)}},  \label{2.3a}
\end{equation}
with respect to $g_{ab}$ yields the field equations,
\begin{eqnarray}
G_{ab}^{(5)} &=&\kappa _5^2T_{ab}^{(5,\;\phi )}  \label{2.3b} \\
&&\ +\kappa _5^2\sum_{I=1}^2{{\mathcal{T}}_{\mu \nu }^{(I)}}e_a^{(\mu
)}e_b^{(\nu )}\sqrt{\left| \frac{g^{(I)}}g\right| }\;\delta \left( \Phi
_I\right) ,  \nonumber
\end{eqnarray}
where $T_{ab}^{(5,\;\phi )}$ and ${\mathcal{T}}_{\mu \nu }^{(I)}$'s are the
energy-momentum tensors of the bulk and branes, respectively, and are given
by
\begin{eqnarray}
\label{2.3e}
\kappa _5^2T_{ab}^{(5,\;\phi )} &\equiv &\frac 12\left( \nabla _a\phi
\right) \left( \nabla _b\phi \right)   \nb\\
&&\ -\frac 14g_{ab}\left[ \left( \nabla \phi \right) ^2-12\alpha ^2e^{-2\phi
}\right] ,   \\
\label{2.3ea}
{\mathcal{T}}_{\mu \nu }^{(I)} &\equiv &\left(\tau _\phi ^{(I)} +
g^{(I)}_{k}\right)g_{\mu \nu }^{(I)} + {\tau}_{\mu \nu }^{(I)}, \\
\label{2.3eb}
{{\tau}}_{\mu \nu }^{(I)} &\equiv &2\frac{\delta {\mathcal{L}}%
_{4,\;m}^{(I)}}{\delta g^{(I)\;\mu \nu }}-g_{\mu \nu }^{(I)}{\mathcal{L}}%
_{4,\;m}^{(I)}, \\
\label{2.3ec}
\tau _\phi ^{(I)} &\equiv &6\epsilon _I\alpha \kappa _5^{-2}e^{-\phi }
+ V^{(I)}_{4}(\phi), \\
\label{2.3ed}
e_{(\mu)}^{(I) a} &\equiv &\frac{\partial x^a}{\partial \xi ^\mu_{(I)}}, \\
\label{2.3ee}
g_{\mu \nu }^{(I)} &\equiv &\left. e_{(\mu)}^{(I)\; a}
e_{(\nu)}^{(I)\; b}g_{ab}\right|
_{M^{(I)}_{4}},
\end{eqnarray}
where $\xi^\mu_{(I)} \;(\mu =0,1,2,3)$ are the intrinsic coordinates on the
orbifold branes. $\delta \left( \Phi _I\right) $ denotes the Dirac delta
function, normalized in the sense of \cite{LMW01}. The two orbifold branes
are located on the hypersurfaces,
\begin{equation}
\label{2.3f}
\Phi _I\left( x^a\right) =0,\;(I=1,2),  
\end{equation}
from which we find that the normal vector to the I-th brane is given by
\bq
\label{3.21}
n^{(I)}_a  = \frac 1{N^{(I)}}\frac{\partial \Phi _I(x)}{\partial x^a},  
\eq
where
\bq
\label{3.21aa} 
N^{(I)} \equiv \sqrt{\left| \Phi _{I,c}\Phi _I^{,c}\right|}.
\eq
It is interesting to note that the contribution of the modulus field to the
branes acts as a varying cosmological constant, as can be seen clearly from
Eqs. (\ref{2.3ea}) and (\ref{2.3ec}).

Variation of the total action (\ref{2.3a}) with respect to $\phi $, on the
other hand, yields the generalized Klein-Gordon equation,
\begin{eqnarray}
\label{2.3g}
\Box {\phi } &=&12\alpha ^2e^{-2\phi }+\sum_{I=1}^2{\left({12\alpha
\epsilon _I}e^{-\phi }- 2\kappa^{2}_{5}\frac{\partial V^{(I)}_{4}}{\partial \phi}\right.}
  \nb\\
&& \left. -\sigma _\phi ^{(I)}\right) \sqrt{\left| \frac{g^{(I)}}g\right| }
\;\delta \left(\Phi _I\right),
\end{eqnarray}
where $\Box \equiv g^{ab}\nabla _a\nabla _b$, and
\begin{equation}
\sigma _\phi ^{(I)}\equiv -2\kappa _5^2\;\frac{\delta {\mathcal{L}}%
_{4,\;m}^{(I)}}{\delta \phi }.  \label{2.3h}
\end{equation}
Note the difference signs of $\sigma _\phi ^{(I)}$ defined here and the one
used in \cite{GWW07}.

To solve Eqs. (\ref{2.3b}) and (\ref{2.3g}), it is
found convenient to separate them into two groups: one is defined
outside the two orbifold branes, and the other is defined on the two
branes.

\subsection{Field Equations Outside the Two Branes}

To obtain the equations outside the two orbifold branes is straightforward,
and they are simply the 5-dimensional Einstein field equations (\ref{2.3b}),
and the matter field equation Eq. (\ref{2.3g}) without the delta function
parts,
\begin{eqnarray}
G_{ab}^{(5)} &=&\frac 12\left( \nabla _a\phi \right) \left( \nabla _b\phi
\right)  \label{3.1a} \\
&&\ -\frac 14g_{ab}\left[ \left( \nabla \phi \right) ^2-12\alpha ^2e^{-2\phi
}\right] ,  \nonumber \\
\Box \phi &=&12\alpha ^2e^{-2\phi }.  \label{3.1b}
\end{eqnarray}
Therefore, in the rest of this section, we shall concentrate ourselves on
the derivation of the field equations on the branes.

\subsection{Field Equations on the Two Orbifold Branes}

To obtain the field equations on the two orbifold branes, one can follow two
different approaches: (1) First express the delta function parts in the
left-hand sides of Eqs. (\ref{2.3b}) and (\ref{2.3g}) in terms of the
discontinuities of the first derivatives of the metric coefficients and
matter fields, and then equal the corresponding delta function parts in the
right-hand sides of these equations, as shown systematically in \cite{WCS08}%
. (2) The second approach is to use the Gauss-Codacci and Lanczos equations
to write down the $4$-dimensional gravitational field equations on the
branes \cite{SMS,AG04}. It should be noted that these two approaches are
equivalent and complementary one to the other. In this paper, we follow the
second approach to obtain the gravitational field equations, and the first 
approach to obtain the matter field equations on the
two branes.

\subsubsection{Gravitational Field Equations on the Two Branes}

For a timelike brane, the 4-dimensional Einstein tensor $G_{\mu \nu }^{(4)}$
can be written as \cite{SMS,AG04,WS09},
\begin{equation}
G_{\mu \nu }^{(4)}={\mathcal{G}}_{\mu \nu }^{(5)}+E_{\mu \nu }^{(5)}+{%
\mathcal{F}}_{\mu \nu }^{(4)},  \label{3.2}
\end{equation}
with
\begin{eqnarray}
{\mathcal{G}}_{\mu \nu }^{(5)} &\equiv &\frac 23\left\{ G_{ab}^{(5)}e_{(\mu
)}^ae_{(\nu )}^b\right.  \label{3.3} \\
&&\left. -\left[ G_{ab}n^an^b+\frac 14G^{(5)}\right] g_{\mu \nu }\right\} ,
\nonumber \\
E_{\mu \nu }^{(5)} &\equiv &C_{abcd}^{(5)}n^ae_{(\mu )}^bn^ce_{(\nu )}^d,
\nonumber \\
{\mathcal{F}}_{\mu \nu }^{(4)} &\equiv &K_{\mu \lambda }K_\nu ^\lambda
-KK_{\mu \nu }  \nonumber \\
&&-\frac 12g_{\mu \nu }\left( K_{\alpha \beta }K^{\alpha \beta }-K^2\right) ,
\nonumber
\end{eqnarray}
where 
$G^{(5)}\equiv
g^{ab}G_{ab}^{(5)}$, and $C_{abcd}^{(5)}$ the Weyl tensor. The extrinsic
curvature $K_{\mu \nu }$ is defined as
\begin{equation}
K_{\mu \nu }\equiv e_{(\mu )}^ae_{(\nu )}^b\nabla _an_b.  \label{3.4a}
\end{equation}
A crucial step of this approach is the Lanczos equations \cite{Lan22},
\begin{equation}
\left[ K_{\mu \nu }^{(I)}\right] ^{-}-g_{\mu \nu }^{(I)}\left[
K^{(I)}\right] ^{-}=-\kappa _5^2{\mathcal{T}}_{\mu \nu }^{(I)},  \label{3.4b}
\end{equation}
where
\begin{eqnarray}
\left[ K_{\mu \nu }^{(I)}\right] ^{-} &\equiv &\mathrm{lim}_{\Phi
_I\rightarrow 0^{+}}K_{\mu \nu }^{(I)\;+}-\mathrm{lim}_{\Phi _I\rightarrow
0^{-}}K_{\mu \nu }^{(I)\;-},  \nonumber  \label{3.5} \\
\left[ K^{(I)}\right] ^{-} &\equiv &g^{(I)\;\mu \nu }\left[ K_{\mu \nu
}^{(I)}\right] ^{-}.
\end{eqnarray}
On the other hand, from the Codacci equation, one finds \cite{AG04,WS09}
\bq
\lb{2.3gaa}
G^{(5)}_{ab}n^{(I) a}e^{(I)b}_{(\mu)}  = 
\left(K^{(I)\; \mu}_{\nu} - \delta^{\mu}_{\nu}K^{(I)}\right)_{;\mu},
\eq
where a semicolon ``;" denotes the covariant derivative with respect to the
reduced metric $g^{(I)}_{\mu\nu}$. The combination of Eqs. (\ref{3.4b}) 
and (\ref{2.3gaa}) yields the conservation law,
\bq
\lb{2.3ga}
\left[G^{(5)}_{ab}n^{(I) a}e^{(I)b}_{(\mu)}\right]^{-} = - 
\kappa^{2}_{5}{\cal{T}}^{(I)\lambda}_{\;\;\;\;\;\; \mu;\lambda}.
\eq
Since $n^{(I) a} e^{(I)\; b}_{(\mu)} g_{ab} = 0$, from Eqs. (\ref{2.3b}),
(\ref{2.3e}), and (\ref{2.3ga}), we find 
 \bq
\lb{2.3gc}
{\cal{T}}^{(I)\lambda}_{\;\;\;\;\;\; \mu;\lambda} =  - \frac{1}{2\kappa^{2}_{5}}
\left[\phi_{,n} \phi_{,\mu}\right]^{-},
\eq
where $\phi_{,n} \equiv n^{ a}\phi_{,a}$ and 
$\phi_{,\mu} \equiv e^{ a}_{(\mu)}\phi_{,a}$.

Assuming that the branes have $Z_{2}$ symmetry, we have
\begin{equation}  \label{3.6}
K^{(I)\; +}_{\mu\nu} = - K^{(I)\; -}_{\mu\nu}.
\end{equation}
Then, we can express the intrinsic curvatures $K^{(I)}_{\mu\nu}$ appearing
in the expression of ${\mathcal{F}}^{(4)}_{\mu\nu}$ in terms of the
effective energy-momentum tensor ${\mathcal{T}}_{\mu\nu} ^{(I)}$ through the
Lanczos equations (\ref{3.4b}).
Hence, $G^{(4)}_{\mu\nu}$ given by Eq. (\ref{3.2}) can be cast in the form
\cite{WS07},
\begin{eqnarray}  \label{3.9}
G^{(4)}_{\mu\nu} &=& {\mathcal{G}}^{(5)}_{\mu\nu} + E^{(5)}_{\mu\nu} + {%
\mathcal{E}}_{\mu\nu}^{(4)} + \kappa^{4}_{5}\pi_{\mu\nu}  \nonumber \\
& & + \kappa^{2}_{4}\tau_{\mu\nu} + \Lambda_{4} g_{\mu\nu},
\end{eqnarray}
where
\begin{eqnarray}  \label{3.10}
\pi_{\mu\nu} &\equiv& \frac{1}{4}\left\{\tau_{\mu\lambda}\tau^{\lambda}_{%
\nu} - \frac{1}{3}\tau \tau_{\mu\nu} - \frac{1}{2}g_{\mu\nu}\left(\tau^{%
\alpha\beta} \tau_{\alpha\beta} - \frac{1}{3}\tau^{2}\right)\right\},
\nonumber \\
{\mathcal{E}}_{\mu\nu}^{(4)} &\equiv& \frac{\kappa^{4}_{5}}{6}\tau_{\phi}
\left[\tau_{\mu\nu} + \left(g_{k} + \frac{1}{2}\tau_{\phi}\right)g_{\mu%
\nu}\right],
\end{eqnarray}
and
\begin{equation}  \label{3.11}
\kappa^{2}_{4} = \frac{1}{6}g_{k}\kappa^{4}_{5},\;\; \Lambda_{4} = \frac{1%
}{12}g_{k}^{2}\kappa^{4}_{5}.
\end{equation}
It should be noted that in writing the above equations, we implicitly assumed 
that only the brane tension $g_{k}$ couples with the 4-dimensional Newtonian
constant $G$ through Eq.(\ref{3.11}), where $\kappa^{2}_{4} = 8\pi G/c^{4}$. 
This is in the same spirit as first proposed in \cite{Cline99}. However, in the 
literature \cite{branes}, some argued that other matter fields, including
the scalar potential $V_{4}(\phi)$, should also contribute to $G$. In the 
latter case, one can see that the resulted 4D Newtonian constant is model-dependent, 
and in general a function of time and space $G = G\left(t, x^{i}\right)$. In the 
former case,  the 4D Newtonian constant will be uniquely determined 
once the tension of the brane is given.  However, this does not mean that 
the former has no problem at all. In particular, considering the fact that in 
the original derivation of the action (\ref{2.2}), the tension of the branes 
was not included \cite{LOSW99}, one might argue that such an assumption is 
problematic, too. While this is indeed a very subtle  problem, in this paper 
we shall take the point of view of \cite{Cline99}, and assume that only brane 
tension is related to $G$.

For a perfect fluid,
\begin{equation}  \label{3.12}
\tau_{\mu\nu} = \left(\rho + p\right)u_{\mu}u_{\nu} - p g_{\mu\nu},
\end{equation}
where $u_{\mu}$ is the four-velocity of the fluid on the brane, we find that
\begin{equation}  \label{3.13}
\pi_{\mu\nu} = \frac{1}{6}\rho \left[\left(\rho + p\right)u_{\mu}u_{\nu} -
\left(p + \frac{1}{2}\rho\right)g_{\mu\nu}\right].
\end{equation}
Note that in writing Eqs. (\ref{3.9})-(\ref{3.13}), without causing any
confusion, we had dropped the super indices $(I)$.

It should also be noted that the definitions of $\kappa_{4}$ and $%
\Lambda_{4} $ in Eq. (\ref{3.11}) are unique, because in Eqs. (\ref{3.9})
their corresponding terms are the only ones that linearly proportional to
the matter field $\tau_{\mu\nu}$ and the spacetime geometry $g_{\mu\nu}$. In
addition, they are exactly the ones widely used in brane-worlds \cite{branes}.

\subsubsection{Matter Field Equations on the Two Branes}

\begin{figure}
\centering
\includegraphics[width=8cm]{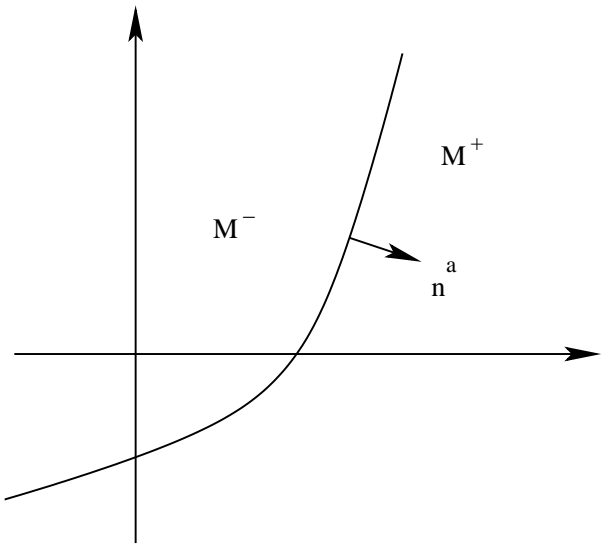}
\caption{The  surface $\Phi_{I}(x) = 0$ divides the spacetimes into two regions, 
$\Phi_{I}(x) > 0$ 
and $\Phi_{I}(x) < 0$. The normal vector defined by Eq.(\ref{3.21}) points from $M^{-}$ to
$M^{+}$, where $M^{+} \equiv \left\{x: \Phi_{I}(x) > 0\right\}$ and
$M^{-} \equiv \left\{x: \Phi_{I}(x) < 0\right\}$. }
\label{fig0}
\end{figure}

On the other hand, the I-th brane, localized on the surface $\Phi _I(x)=0$,
divides the spacetime into two regions, one with $\Phi _I(x)>0$ and the
other with $\Phi _I(x)<0$ [Cf. Fig. \ref{fig0}]. Since the field equations 
are the second-order
differential equations, the matter fields have to be at least continuous
across this surface, although in general their first-order derivatives  are
not. Introducing the Heaviside function, defined as
\begin{equation}
H\left( x\right) =\left\{ \matrix{1, & x > 0,\cr 0, & x < 0,\cr}\right.
\label{3.14}
\end{equation}
for any given $C^0$ function $F(x)$,  in the neighborhood of $\Phi _I(x)=0$
 we can always write it in the  form,
\begin{equation}
F(x)=F^{+}(x)H\left( \Phi _I\right) +F^{-}(x)\left[ 1-H\left( \Phi _I\right)
\right] ,  \label{3.15}
\end{equation}
where $F^{+}\;(F^{-})$ is defined in the region $\Phi _I>0\;(\Phi _I<0)$,
and
\begin{equation}
\left. F^{+}(x)\right| _{\Phi _I=0^{+}}=\left. F^{-}(x)\right| _{\Phi
_I=0^{-}}.  \label{3.15a}
\end{equation}
Then, we find that
\begin{eqnarray}
F_{,a}(x) &=&F_{,a}^{+}(x)H\left( \Phi _I\right) +F_{,a}^{-}(x)\left[
1-H\left( \Phi _I\right) \right] ,  \nonumber  \label{3.16} \\
F_{,ab}(x) &=&F_{,ab}^{+}(x)H\left( \Phi _I\right) +F_{,ab}^{-}(x)\left[
1-H\left( \Phi _I\right) \right]  \nonumber \\
&&+\left[ F_{,a}\right] ^{-}\frac{\partial \Phi _I(x)}{\partial x^b}\;\delta
\left( \Phi _I\right) ,
\end{eqnarray}
where $\left[ F_{,a}\right] ^{-}$ is defined as that in Eq. (\ref{3.5}).
Projecting $F_{,a}$ onto $n^a$ and $e_{(\mu )}^a$ directions, we find
\begin{equation}
F_{,a}=F_{,\mu }e_a^{(\mu )}-F_{,n}n_a,  \label{3.17}
\end{equation}
where
\begin{equation}
F_{,n}\equiv n^aF_{,a},\;\;F_{,\mu }\equiv e_{(\mu )}^aF_{,a}.  \label{3.18}
\end{equation}
Then, it can be shown that
\begin{eqnarray}
&&\left[ F_{,n}\right] ^{-}=\left[ F_{,a}\right] ^{-}n^a\not =0,  \nonumber
\label{3.19} \\
&&\left[ F_{,\mu }\right] ^{-}=\left[ F_{,a}\right] ^{-}e_{(\mu )}^a=0.
\end{eqnarray}
Inserting Eqs. (\ref{3.17})-(\ref{3.19}) into Eq. (\ref{3.16}), we find
\begin{eqnarray}
F_{,ab}(x) &=&F_{,ab}^{+}(x)H\left( \Phi _I\right) +F_{,ab}^{-}(x)\left[
1-H\left( \Phi _I\right) \right]  \nonumber  \label{3.20} \\
&&-\left[ F_{,n}\right] ^{-}n^{(I)}_an^{(I)}_bN^{(I)}\;\delta \left( \Phi _I\right).
\end{eqnarray}
Due to the $Z_{2}$ symmetry, we can further write $\left[ F^{(I)}_{,n}\right] ^{-}$
as
\bq
\lb{3.21a}
\left[ F^{(I)}_{,n}\right] ^{-} = - 2\epsilon_{I}F^{(I)}_{,n},
\eq
where
\bqn
\lb{3.21b}
F^{(1)}_{,n} &\equiv& \lim_{\Phi_{1} \rightarrow 0^{-}}{\left(n^{a}F_{,a}\right)}\nb\\
F^{(2)}_{,n} &\equiv& \lim_{\Phi_{2} \rightarrow 0^{+}}{\left(n^{a}F_{,a}\right)}.
\eqn
Substituting Eq. (\ref{3.20}) into Eq. (\ref{2.3g}), we find that the matter
field equation on the branes reads,
\begin{eqnarray}
 \label{3.22}
\phi _{,n}^{(I)} &=&\frac{\epsilon_{I}}{2N^{(I)}}\left(2\kappa^{2}_{5}
\frac{\partial V^{(I)}_{4}}{\partial \phi} - 12\alpha
\epsilon _Ie^{-\phi } \right.\nb\\
&&\left. + \sigma _\phi ^{(I)}\right) \sqrt{\left| \frac{g^{(I)}}g\right|},
\end{eqnarray}
where $\phi _{,n}^{(I)}$ is defined as that given by Eq. (\ref{3.21b}).
Similarly, Eq. (\ref{2.3gc}) can be written as
\bq
\lb{2.3gca}
{\cal{T}}^{(I)\lambda}_{\;\;\;\;\;\; \mu;\lambda} =  
\frac{\epsilon_{I}}{\kappa^{2}_{5}}\phi^{(I)}_{,n} \phi^{(I)}_{,\mu}.
\eq

Eqs. (\ref{3.1a}), (\ref{3.1b}), (\ref{3.9}), (\ref{3.22}), and (\ref{2.3gca}) 
consist of the complete set of both the gravitational and the matter field 
equations in the framework of the Horava-Witten heterotic M-Theory on $S^{1}/Z_{2}$.

\section{ Radion stability and radion mass}

\renewcommand{\theequation}{3.\arabic{equation}} \setcounter{equation}{0}

In the studies of orbifold branes, an important issue is the radion
stability \cite{branes}. In this section, we shall address this problem.

\subsection{Static Solution with 4D Poincar\'e Symmetry}

\begin{figure}[tbp]
\includegraphics[width=\columnwidth]{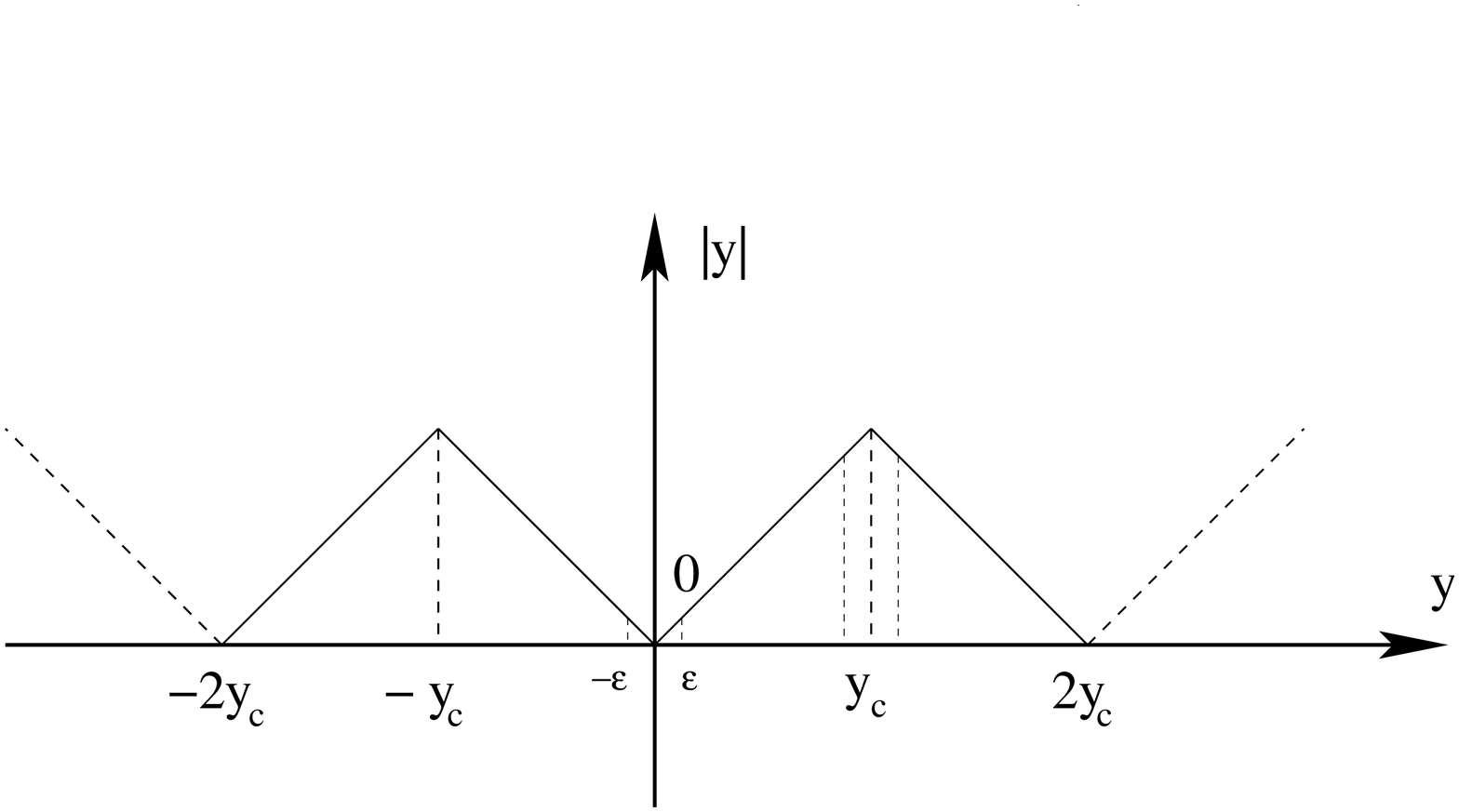}
\caption{The function $\left|y\right|$ appearing in Eq. (\ref{5.1ba}). }
\label{fig1}
\end{figure}

To begin with,  let us first consider the 5-dimensional static
metric with a 4-dimensional Poincar\'e symmetry  \cite{LOSW99},
\bq
\label{5.1b}
ds_5^2 = e^{2\sigma (y)}\left( \eta _{\mu \nu }dx^\mu dx^\nu -dy^2\right),
\eq
where
\bqn
\label{5.1ba}
\sigma (y) &=&\frac 15\ln \left( \frac{|y|+y_0}L\right) ,  \nonumber \\
\phi (y) &=&\frac 65\ln \left( \frac{|y|+y_0}L\right) +\phi_0,  \nonumber \\
\phi_0 &=& \ln \left(5L\alpha \right)  \nonumber
\end{eqnarray}
where $|y|$ is defined as that given in Fig.\ref{fig1}, ${L}$ and $y_0$ are
positive constants.

Then, it can be shown that the above solution satisfies the gravitational
and matter field equations outside the branes, Eqs. (\ref{3.1a}) and (\ref{3.1b}).
On the two branes, assuming that the spacetime is vacuum, i.e., $\tau _{\mu
\nu }^{(I)}=0 = \sigma^{(I)}_{\phi}$,  Eqs. (\ref{3.9}) and (\ref{3.22}) require
\bqn
\label{5.2}
& & V^{(I)}_{4}\left(\phi_{I}\right) + g^{(I)}_{k} = 0,\\  
\label{5.2aa}
& & \frac{\partial V^{(I)}_4\left(\phi_{I}\right)}{\partial \phi} = 0,
\eqn
while Eq. (\ref{2.3gca}) is satisfied identically,
where $\phi_{I} \equiv \left.\phi\right|_{y = y_{I}}$.

To study the 4-dimensional effective gravitational coupling, as well as the
radion stability, it is found convenient to introduce the proper distance $Y$,
defined by
\begin{equation}
Y=\left( \frac{5L}6\right) \left\{ \left( \frac{y+y_0}L\right) ^{6/5}-\left(
\frac{y_0}L\right) ^{6/5}\right\} .  \label{5.2a}
\end{equation}
Then, in terms of $Y$, the static solution (\ref{5.1b}) can be written as
\begin{equation}
ds_5^2=e^{-2A(Y)}\eta _{\mu \nu }dx^\mu dx^\nu -dY^2,  \label{5.1aa}
\end{equation}
with
\begin{eqnarray}
A(Y) &=&-\frac 16\ln \left\{ \left( \frac 6{5L}\right) \left( |Y|+Y_0\right)
\right\} , \label{5.1aaa}\\
\phi (Y) &=&\ln \left\{ \left( \frac 6{5L}\right) \left( |Y|+Y_0\right)
\right\} +\phi _0, \label{5.1ab}
\end{eqnarray}
where $|Y|$ is defined also as that of Fig. \ref{fig1}, with
\begin{eqnarray}
Y_0 &\equiv &\left( \frac{5L}6\right) \left( \frac{y_0}L\right) ^{6/5},
\nonumber  \label{5.2b} \\
Y_c &\equiv &\left( \frac{5L}6\right) \left\{ \left( \frac{y_c+y_0}L\right)
^{6/5}-\left( \frac{y_0}L\right) ^{6/5}\right\} ,
\end{eqnarray}
and $Y_2=0,\;Y_1=Y_c$.

\subsection{Radion Stability}

Following \cite{GW99}, let us consider a massive scalar field $\Phi $ with
the actions,
\begin{eqnarray}
S_b &=&\int {d^4x}\int_0^{Y_c}{dY\sqrt{-g_5}\left( \left( \nabla \Phi
\right) ^2- M^2\Phi ^2\right) },  \nonumber  \label{5.3a} \\
S_I &=&-\alpha _I\int_{M_4^{(I)}}{d^4x\sqrt{-g_4^{(I)}}\left( \Phi
^2-v_I^2\right) ^2},
\end{eqnarray}
where $\alpha _I$ and $v_I$ are real constants. Then, it can be shown that,
in the background of Eq. (\ref{5.1aa}), the massive scalar field $\Phi $
satisfies the following Klein-Gordon equation
\begin{equation}
\Phi ^{\prime \prime }-4A^{\prime }\Phi ^{\prime }-M^2\Phi =\sum_{I=1}^2{%
2\alpha _I\Phi \left( \Phi ^2-v_I^2\right) \delta (Y-Y_I)},  \label{5.5}
\end{equation}
where a prime denotes the ordinary derivative with respect to the indicated
argument, which in the present case is $Y$. 
Integrating the above equation in the neighborhood of the I-th brane, we
find that
\begin{equation}
\left. \frac{d\Phi (Y)}{dY}\right| _{Y_I-\epsilon }^{Y_I+\epsilon }=2\alpha
_I\Phi _I\left( \Phi _I^2-v_I^2\right) ,  \label{5.5a}
\end{equation}
where $\Phi _I\equiv \Phi (Y_I)$. Setting
\begin{equation}
z\equiv M(Y+Y_0), \;\;\;
\Phi =\left(\frac zM\right)^{1/6}u(z),  \label{5.5b}
\end{equation}
we find that, outside of the branes, Eq. (\ref{5.5}) reduces,
\begin{equation}
\frac{d^2u}{dz^2}+\frac 1z\frac{du}{dz}-\left(1+\frac{\nu ^2}{z^2}\right)u=0,
\label{5.5c}
\end{equation}
where $\nu \equiv 1/6$. Eq. (\ref{5.5c}) is the standard modified Bessel
equation \cite{AS72}, which has the general solution
\begin{equation}
u(z)=aI_\nu (z)+bK_\nu (z),  \label{5.5d}
\end{equation}
where $I_\nu (z)$ and $K_\nu (z)$ denote the modified Bessel functions, and $%
a$ and $b$ are the integration constants, which are uniquely determined by
the boundary conditions (\ref{5.5a}). Since
\begin{eqnarray}
\lim_{Y\rightarrow Y_c^{+}}{\frac{d\Phi (Y)}{dY}} &=&-\lim_{Y\rightarrow
Y_c^{-}}{\frac{d\Phi (Y)}{dY}}\equiv -\Phi ^{\prime }\left( Y_c\right) ,
\nonumber  \label{5.5e} \\
\lim_{Y\rightarrow 0^{-}}{\frac{d\Phi (Y)}{dY}} &=&-\lim_{Y\rightarrow 0^{+}}%
{\frac{d\Phi (Y)}{dY}}\equiv - \Phi ^{\prime }(0),
\end{eqnarray}
we find that the conditions (\ref{5.5a}) can be written in the forms,
\begin{eqnarray}
\label{5.5fa}
\Phi ^{\prime }(Y_c) &=&-\alpha _1\Phi _1\left( \Phi _1^2-v_1^2\right), \\
\label{5.5fb}
\Phi ^{\prime }(0) &=&\alpha _2\Phi _2\left( \Phi _2^2-v_2^2\right).
\end{eqnarray}

Inserting the above solution back to the actions (\ref{5.3a}), and then
integrating them with respect to $Y$, we obtain the effective potential for
the radion $Y_{c}$,
\begin{eqnarray}  \label{5.5g}
V_{\Phi}\left(Y_{c}\right) &\equiv& - \int_{0+\epsilon}^{Y_{c}- \epsilon}{dY
\sqrt{-g_{5}} \left(\left(\nabla\Phi\right)^{2} - M^{2}\Phi^{2}\right)}
\nonumber \\
& & + \sum_{I=1}^{2}{\ \alpha_{I} \int_{Y_{I} -\epsilon}^{Y_{I} + \epsilon} {%
dY \sqrt{-g_{4}^{(I)}} \left(\Phi^{2} - v^{2}_{I}\right)^{2}}}  \nonumber \\
& & \;\;\;\;\;\;\; \;\;\;\;\;\;\; \times \delta\left(Y-Y_{I}\right)
\nonumber \\
&=& \left. e^{-4A(Y)}\Phi(Y)\Phi^{\prime}(Y)\right|^{Y_{c}}_{0}  \nonumber \\
& & + \sum_{I=1}^{2}{\alpha_{I} \left(\Phi^{2}_{I} -
v^{2}_{I}\right)^{2}e^{-4A(Y_{I})}}.
\end{eqnarray}
In the limit that $\alpha_{I}$'s are very large \cite{GW99}, Eqs. (\ref{5.5fa})
and (\ref{5.5fb}) show that there are solutions only when $\Phi(0) \simeq v_{2}$
and $\Phi(Y_{c}) \simeq v_{1}$, that is,
\begin{eqnarray}
\label{5.5ha}
v_{1} &\simeq& (Y_c+Y_0)^\frac{1}{6}[aI_{\nu}(z_c)+bK_{\nu}(z_c)], \\
\label{5.5hb}
v_{2} &\simeq& Y_0^\frac{1}{6}[aI_{\nu}(z_0)+b_{\nu}(z_0)],
\end{eqnarray}
where $z_{0} \equiv MY_{0}$ and $z_{c} \equiv M(Y_{c} + Y_{0})$. Eqs. (\ref
{5.5ha}) and (\ref{5.5hb}) have the solutions,
\begin{eqnarray}  \label{5.5i}
a &=& \frac{1}{\Delta}\left(K_{\nu}^{(0)}\bar{v_1}-k_{\nu}^{(c)}\bar{v_2}\right),
\nonumber \\
b &=& \frac{1}{\Delta}\left(I_{\nu}^{(c)}\bar{v_2}-I_{\nu}^{(0)}\bar{v_1}\right),
\end{eqnarray}
where $K^{(I)}_{\nu} \equiv K_{\nu}(z_{I}),\; I^{(I)}_{\nu} \equiv
I_{\nu}(z_{I})$, and
\bqn  
\label{5.5j}
\Delta &\equiv& I_{\nu}^{(c)}K_{\nu}^{(0)}-I_{\nu}^{(0)}K_{\nu}^{(c)},\nb\\
\bar{v_1} &=& v_1 \left(\frac{M}{z_c}\right)^{{1}/{6}},\nb\\
\bar{v_2} &=& v_2\left(\frac{M}{z_0}\right)^{{1}/{6}}.
\eqn
Inserting the above expressions into Eq. (\ref{5.5g}), we find that
\bq  
\label{5.5ja}
V_{\Phi}\left(Y_{c}\right) \simeq \left(\frac{6}{5}\right)^{2/3}\left(I\left(z_{c}\right)
- I\left(z_{0}\right)\right),   
\eq
where
\bqn  
\label{5.5jb}
I\left(z\right) &\equiv& a^{2}\left(\nu + z\right)I^{2}_{\nu}(z)
                + 2 ab \nu I_{\nu}(z) K_{\nu}(z)\nb\\
		& &
		+ b^{2}K^{2}_{\nu}(z). 
\eqn

\subsubsection{$M Y_{0} \gg 1$}

\begin{figure}[tbp]
\includegraphics[width=\columnwidth]{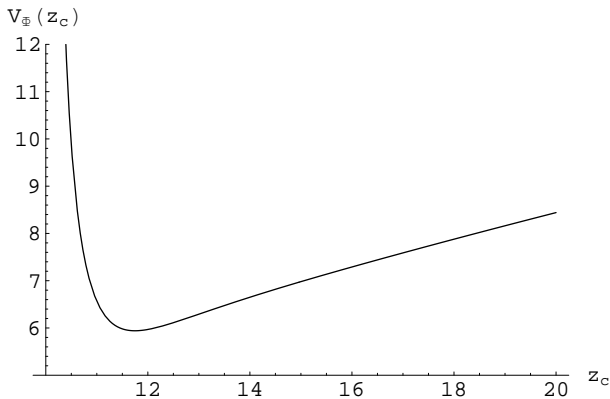}
\caption{The potential defined by Eq. (\ref{5.4}) in the limit of large $v_I$
and $y_0$. In this particular plot, we choose ($z_0$, $v_1$, $v_2$)= (10,
1.0, 0.1).}
\label{fig2}
\end{figure}

\begin{figure}[tbp]
\includegraphics[width=\columnwidth]{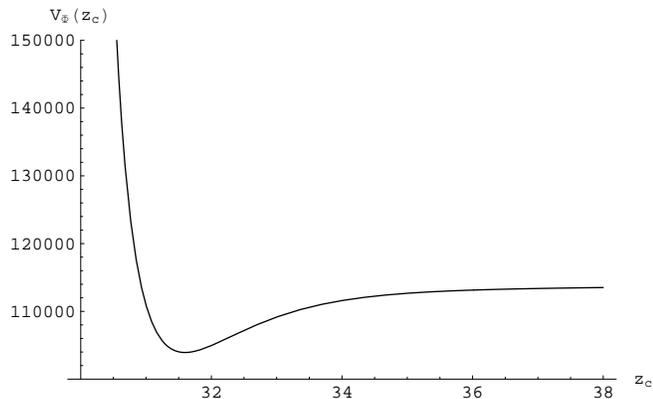}
\caption{The potential defined by Eq. (\ref{5.4}) in the limit of large $v_I$
and $y_0$. In this particular plot, we choose ($z_0$, $v_1$, $v_2$)= (30,
200, 100).}
\label{fig3}
\end{figure}

When $Y_{0} \gg M^{-1}$, we have $z_{0}, \; z_{c} \gg 1$. Then, we find that
\cite{AS72},
\begin{eqnarray}  \label{5.5kc}
I_{\nu}(z) &\simeq& \frac{e^z}{\sqrt{2\pi z}},  \nonumber \\
K_{\nu}(z) &\simeq& \sqrt{\frac{\pi}{2z}}e^{-z},
\end{eqnarray}
for $z\gg 1$.
Substituting them into Eq. (\ref{5.5g}), we find that
\begin{eqnarray}  \label{5.4}
V_{\Phi}\left(Y_{c}\right) &\simeq&  M \left(\frac{6Y_{0}}{5L}\right)^{{2}/{3}}
\left(\left(v_{1}^{2} + v^{2}_{2}\right) \coth\left(z_{c} - z_{0}\right)\right.\nb\\
&& \left. - \frac{2v_{1}v_{2}}{\sinh\left(z_{c} - z_{0}\right)}\right). 
\end{eqnarray}
Thus, we find that
\begin{equation}  \label{5.6}
V_{\Phi}\left(Y_{c}\right) \simeq
V_{\Phi}^{(0)} \times \left\{\matrix{\frac{(v_{1} - v_{2})^{2}z_0^{2/3}}
{\sinh\left(z_{c}-z_{0}\right)} \rightarrow \infty, & z_{c}
\rightarrow z_{0},\cr v_{1}^{2} z_c^{2/3} \rightarrow \infty, &
z_{c} \rightarrow \infty,\cr}\right.
\end{equation}
where $V_{\Phi}^{(0)} \equiv
M^{1/3}\left({6}/{(5L)}\right)^{2/3}$. Figs. \ref{fig2} and
\ref{fig3} show the potential for $(z_{0}, \; v_{1}, \; v_{2}) =
(10, \; 1.0,\;0.1)$ and $(z_{0}, \; v_{1}, \; v_{2}) = (30, \;
200,\;100)$, respectively, from which we can see clearly that it has
a minimal. Therefore, the radion is indeed stable in our current
setup.

\subsubsection{$M Y_{0} \ll 1$}

When $M Y_{0} \ll 1$ and $M Y_{c} \ll 1$, we find that \cite{AS72}
\begin{eqnarray}  \label{5.5kaa}
I_{\nu}(z) &\simeq& \frac{z^{\nu}}{2^{\nu} \Gamma(\nu + 1)},  \nonumber \\
K_{\nu}(z) &\simeq& \frac{2^{\nu - 1} \Gamma(\nu)}{z^{\nu}}.
\end{eqnarray}

Substituting them into Eq. (\ref{5.5g}), we obtain
\begin{equation}  \label{5.4a}
V_{\Phi}\left(Y_{c}\right) \simeq \frac{1}{3} M^{1/3}\left(\frac{6}{5L}%
\right)^{2/3} \frac{\left(v_{1} - v_{2}\right)^{2}}{z_{c}^{2\nu} -
z_{0}^{2\nu}}.
\end{equation}
Clearly, in this limit the potential has no minima, and the corresponding
radion is not stable. Therefore, there exists a minimal mass for the scalar
field $\Phi$, say, $M_{c}$, only when $M > M_{c}$ the corresponding radion
is stable.

It should be noted that, in the Randall-Sundrum setup \cite{RS1}, $Y_{c}$ is
required to be $Y_{c} \simeq 38$ in order to solve the hierarchy problem.
However, in the current setup the hierarchy problem may be solved by using the
ADD mechanism \cite{ADD98}, so such a requirement is not needed here. As a
result, the physical brane is not necessarily placed at $Y = Y_{c}$. Thus,
in our current setup, we can take any of the two branes as the physical one,
in which the standard matter fields are assumed to be present.

\subsection{Radion Mass}

To calculate the radion mass, we need first to find the exact relation between the
radion field $\varphi$ and $Y_{c}$. To this end, let us consider the linear 
perturbations given by \cite{GW00,CGR00},  
\bqn
\lb{5.5m1} 
ds^{2}_{5} &=& e^{-2[A(Y)-F(x)]} \eta_{\mu\nu}
dx^{\mu}dx^{\nu} \nb\\
& & - [1+2F(x)]^2dY^{2}.
\eqn
Then, we find 
\bqn
\lb{5.5m2} 
\delta R_5 &=& \frac{2e^{2(A+F)}}{1+2F}\left[(1-6F)(\nabla F)^2\right.\nb\\
& & \left. +(1+6F)\Box F\right].
\eqn
Thus, we obtain 
\bqn
\lb{5.5m3} 
\delta S &=&\frac{1}{\kappa^2_5}\int
dYdx^4\sqrt{g_5}\delta R_5 \nb\\
&=&\frac{2}{\kappa^2_5}\int_{0}^{Y_{c}}e^{-2A}dY\nb\\
& & \times \int
dx^4e^{-2F}(\nabla F)^2(6F-3).
\eqn
Following \cite{GW00}, by defining
$\varphi=\sqrt{12f}e^{-F}\sqrt{1-2F}$, we obtain
\begin{equation}
\label{5.5m4}
    \delta S=-\frac12\int dx^4(\nabla \varphi)^2,
\end{equation}
where
\begin{eqnarray}
\label{5.5m5}
    f=\frac{1}{k^2_5}\int_{0}^{Y_{c}}e^{-2A}dY.
\end{eqnarray}
Substituting Eq. (\ref {5.1aaa}) into
Eq. (\ref{5.5m5}), and in the limit $F(x)\rightarrow 0$, we can write
 $\varphi$ as
\begin{eqnarray}\label{5.5m6}
    \varphi(Y_c)&=&3\sqrt{2}\left(\frac 65\right) ^{1/6} M_5^{3/2}L^{1/2}\nb\\
    && \times \left\{ \left( \frac{Y_c+Y_0}{{L}}\right) ^{4/3}-\left( \frac{Y_0%
}{{L}}\right) ^{4/3}\right\}^{1/2}, \;\;\;\;
\end{eqnarray}
where $M^3_5=\kappa^{-2}_5$, as can be seen from Eqs. (\ref{3.11}).
When $z_0=MY_0\gg1$, the potential $V(Y_c)$ given by Eq. (\ref{5.4}) has a
minimum at
\begin{equation}
\label{5.5m7}
    MY_c=z_c-z_0=\ln\left(\frac{v_1}{v_2}\right),
\end{equation}
where, without loss of generality, we has set $v_1>v_2$.

Combining Eq. (\ref{5.4}) with Eq. (\ref{5.5m6}), we obtain the mass of
$\varphi$, which in the large $MY_0$ limit is given by
\bqn
\label{5.5m8}
    m_{\varphi}&=& \sqrt\frac{\partial^2V}{2\partial\varphi^2}\approx
    M^{-1/2}\frac{v^2_1}{v_2}\left(\frac{Y_0}{L}\right)^{1/6}\nb\\
    & & \times 
    \sqrt\frac{\ln\left({v_1}/{v_2}\right)}
    {\left(\left(\frac{v_1}{v_2}\right)^2-1\right)^3}.
\eqn
Note that $v_{1,2}$ have the dimension of $[m]^{3/2}$ \cite{GW99}. Then, without loss 
of generality, we assume that $v_{i} = M^{3/2} \tilde{v}_{i}$, where $\tilde{v}_{i}$ 
should be  order of one.  For such a choice, the last factor in the right-hand side
of Eq.(\ref{5.5m8}) is also order of one. Without introducing a new hierarchy, we
would also expect that $\left({Y_0}/{L}\right)^{1/6} \simeq {\cal{O}}(1)$. On the other hand,
since $M$ is the mass of the bulk scalar field, we would expect that $M \simeq M_{5}$, where
$M_{5}$ is the 5-dimensional Planck mass, given by $M_{5} = M^{3}_{11}R^{2}$, where $R$
is the typical size of the extra dimensions \cite{GWW07}. To have the effective cosmological
constant be in the order of observation,  $\rho_{\Lambda} \simeq 10^{-47} \; GeV^{4}$, it was
found that $R \simeq 10^{-22}\; m$ for $M_{11} \simeq TeV$ \cite{GWW07}. Putting all
 these arguments together, we find that
\bqn
\label{5.5m9}
    m_{\varphi} &\simeq& M \simeq M_{5} =\left(\frac{M_{11}}{M_{pl}}\right)^{3} 
    \left(\frac{R }{l_{pl}}\right)^{2} M_{pl} \nb\\
    &\simeq& 0.1\; GeV,
\eqn
which is much higher than the current observational limit $m_{\varphi} \ge 10^{-3}\; eV$
\cite{RS1}.


\section{ Localization of Gravity and 4D Effective Newtonian Potential}

\renewcommand{\theequation}{4.\arabic{equation}} \setcounter{equation}{0}

To study the localization of gravity and the four-dimensional effective
gravitational potential, in this section let us consider small fluctuations $%
h_{ab}$ of the 5-dimensional static metric with a 4-dimensional Poincar\'e
symmetry, given by Eq. (\ref{5.1b}) in its conformally flat form.

\subsection{Tensor Perturbations and the KK Towers}

\begin{figure}[tbp]
\centering
\includegraphics[width=8cm]{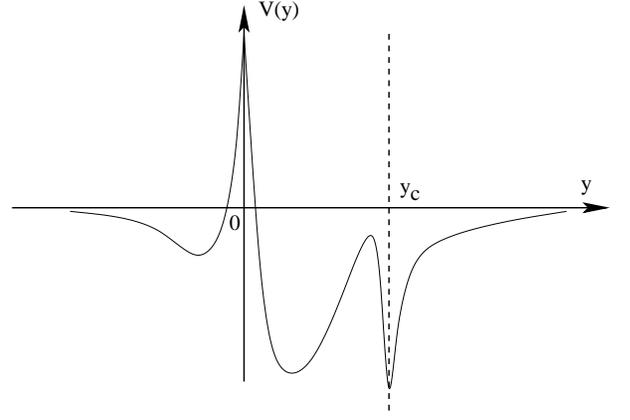}
\caption{The potential defined by Eq. (\ref{7.7}). }
\label{fig5}
\end{figure}

Since such tensor perturbations are not coupled with scalar ones \cite{GT00}%
, without loss of generality, we can set the perturbations of the scalar
field $\phi$  to zero, i.e., $\delta \phi =0$. We
shall choose the gauge \cite{RS2},
\begin{equation}
h_{ay}=0,\;\;\;h_\lambda ^\lambda =0=\partial ^\lambda h_{\mu \lambda }.
\label{7.1}
\end{equation}
Then, it can be shown that \cite{Csaki00}
\begin{eqnarray}
\delta {G}_{ab}^{(5)} &=&-\frac 12\Box _5h_{ab}-\frac 32\left\{ \left(
\partial _c\sigma \right) \left( \partial ^ch_{ab}\right) \right.  \nonumber
\label{7.2} \\
&&\left. -2\left[ \Box _5\sigma +\left( \partial _c\sigma \right) \left(
\partial ^c\sigma \right) \right] h_{ab}\right\} ,  \nonumber \\
\kappa _5^2\delta {T}_{ab}^{(5)} &=&\frac 14\left( {\phi ^{\prime }}%
^2+2e^{2\sigma }V_5\right) h_{ab},  \nonumber \\
\delta {T}_{\mu \nu }^{(4)} &=&\left( \tau _{(\phi ,\psi )}^{(I)}+2\rho
_\Lambda ^{(I)}\right) e^{2\sigma (y_I)}h_{\mu \nu }(x,y_I),
\end{eqnarray}
where $\Box _5\equiv\eta ^{ab}\partial _a\partial _b$ and $\left( \partial
_c\sigma \right) \left( \partial ^ch_{ab}\right)\equiv \eta ^{cd}\left(
\partial _c\sigma \right) \left( \partial _dh_{ab}\right) $, with $\eta
^{ab} $ being the five-dimensional Minkowski metric. Substituting the above
expressions into the Einstein field Eq. (\ref{2.3b}), 
we find that in the present case there is only one independent equation,
given by
\begin{equation}
\Box _5h_{\mu \nu }+3\left( \partial _c\sigma \right) \left( \partial
^ch_{\mu \nu }\right) =0,  \label{7.3}
\end{equation}
which can be further cast in the form,
\begin{equation}
\Box _5\tilde h_{\mu \nu }+\frac 32\left( \sigma ^{\prime \prime }+\frac
32\sigma ^{\prime }\right) \tilde h_{\mu \nu }=0,  \label{7.4}
\end{equation}
where $h_{\mu \nu }\equiv e^{-3\sigma /2}\tilde h_{\mu \nu }$. Setting
\begin{eqnarray}
&&\tilde h_{\mu \nu }(x,y)=\hat h_{\mu \nu }(x)\psi_{n} (y),  \nonumber
\label{7.5} \\
&&\Box _5=\left( \Box _4 - \nabla _y^2\right) = \left(\eta ^{\mu \nu
}\partial _\mu \partial _\nu - \partial _y^2\right) ,  \nonumber \\
&&\Box _4\hat h_{\mu \nu }(x)= - m_{n}^2\hat h_{\mu \nu }(x),
\end{eqnarray}
we find that Eq. (\ref{7.3}) takes the form of the schr\"odinger equation,
\begin{equation}
\left( -\nabla _y^2+V\right) \psi_{n} =m_{n}^2\psi_{n},  \label{7.6}
\end{equation}
where
\begin{eqnarray}
V &\equiv &\frac 32\left( \sigma ^{\prime \prime }+\frac 32{\sigma ^{\prime }%
}^2\right)  \nonumber  \label{7.7} \\
&=&-\frac{21}{100\left( \left| y\right| +y_0\right) ^2}+\frac{3\delta \left(
y\right) }{5y_0}  \nonumber \\
&&-\frac{3\delta \left( y-y_c\right) }{5\left( y_c+y_0\right) }.
\end{eqnarray}
From the above expression we can see clearly that the potential has a
delta-function well at $y=y_c$, which is responsible for the localization of
the graviton on this brane. In contrast, the potential has a delta-function
barrier at $y=0$, which makes the gravity delocalized on the $y=0$ brane.
Fig. \ref{fig5} shows the potential schematically.

Introducing the operators,
\begin{equation}
Q\equiv \nabla _y-\frac 32\sigma ^{\prime },\;\;\;Q^{\dagger }\equiv -\nabla
_y-\frac 32\sigma ^{\prime },  \label{7.8}
\end{equation}
Eq. (\ref{7.6}) can be written in the form of a supersymmetric quantum
mechanics problem,
\begin{equation}
Q^{\dagger }\cdot Q\psi_{n} =m_{n}^2\psi_{n}.  \label{7.9}
\end{equation}
It should be noted that Eq. (\ref{7.9}) itself does not  guarantee that the
operator $Q^{\dagger }\cdot Q$ is Hermitian, because now it is defined only
on a finite interval, $y\in [0,y_c]$. To ensure its Hermiticity, in addition
to writing the differential equation in the Shr\"odinger form, one also
needs to show that it has Hermitian boundary conditions, which can be
formulated as \cite{Csaki01}
\begin{eqnarray}
{\psi ^{\prime }}_n(0)\psi _m(0) &-&\psi _n(0){\psi ^{\prime }}_m(0)={\psi
^{\prime }}_n\left( y_c\right) \psi _m\left( y_c\right)  \nonumber
\label{7.9a} \\
&&-\psi _n\left( y_c\right) {\psi ^{\prime }}_m\left( y_c\right) ,
\end{eqnarray}
for any two solutions of Eq. (\ref{7.9}). To show that in the present case
this condition is indeed satisfied, let us consider the boundary conditions
at $y=0$ and $y=y_c$. Integration of Eq. (\ref{7.6}) in the neighbourhood of $%
y=0$ and $y=y_c$ yields, respectively, the conditions,
\begin{eqnarray}
\lim_{y\rightarrow y_c^{-}}{\psi ^{\prime }(y)} &=&\frac 3{10\left(
y_c+y_0\right) }\lim_{y\rightarrow y_c^{-}}{\psi (y)},  \label{7.12a} \\
\lim_{y\rightarrow 0^{+}}{\psi ^{\prime }(y)} &=&\frac
3{10y_0}\lim_{y\rightarrow 0^{+}}{\psi (y)}.  \label{7.12b}
\end{eqnarray}
Note that in writing the above equations we had used the $Z_2$ symmetry of
the wave function $\psi_{n}$. Clearly, any solution of Eq. (\ref{7.6}) that
satisfies the above boundary conditions also satisfies Eq. (\ref{7.9a}). That
is, the operator $Q^{\dagger }\cdot Q$ defined by Eq. (\ref{7.8}) is indeed a
positive definite Hermitian operator. Then, by the usual theorems we can see
that all eigenvalues $m_n^2$ are non-negative, and their corresponding wave
functions $\psi _n(y)$ are orthogonal to each other and form a complete
basis. Therefore, {the background is gravitationally stable in our current
setup}.

\subsubsection{Zero Mode}

The four-dimensional gravity is given by the existence of the normalizable
zero mode, for which the corresponding wavefunction is given by
\begin{equation}
\psi _0(y)=N_0\left( \frac{|y|+y_0}L\right) ^{3/10},  \label{7.10}
\end{equation}
where $N_0$ is the normalization factor, defined as
\begin{equation}
N_0\equiv 2\left\{ \frac 52L\left[ \left( \frac{y_c+y_0}L\right)
^{8/5}-\left( \frac{y_0}L\right) ^{8/5}\right] \right\} ^{-1/2}.
\label{7.11}
\end{equation}
Eq. (\ref{7.10}) shows clearly that the wavefunction is increasing as $y$
increases from $0$ to $y_c$. Therefore, the gravity is indeed localized near
the $y=y_c$ brane.

\subsubsection{Non-Zero Modes}

\begin{figure}[tbp]
\centering
\includegraphics[width=8cm]{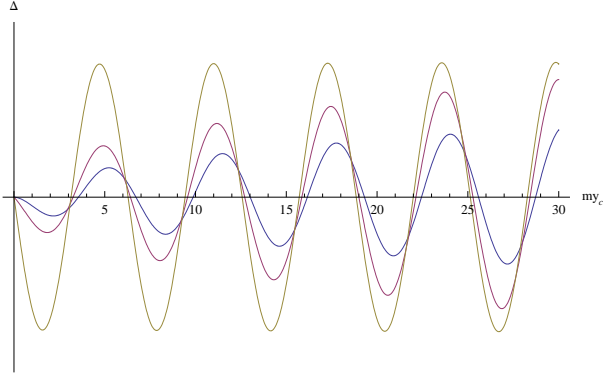}
\caption{The function of $\Delta $ defined by Eq. (\ref{7.18}) for $%
x_0=my_0=0.01,\;1.0,\;1000$, respectively. Note that the horizontal axis is $%
my_c$.}
\label{roots}
\end{figure}

In order to have localized four-dimensional gravity, we require that the
corrections to the Newtonian law from the non-zero modes, the KK modes, of
Eq. (\ref{7.6}), be very small, so that they will not lead to contradiction
with observations. To solve Eq. (\ref{7.6}) outside of the two branes, it is
found convenient to introduce the quantities,
\begin{equation}
\psi (y)\equiv x^{1/2}\;u(x),\;\;\;
 x\equiv m\left( y+y_0\right) .
\label{7.13}
\end{equation}
Then, in terms of $x$ and $u(x)$, Eq.(\ref{7.6}) takes the form,
\begin{equation}
x^2\frac{d^2u}{dx^2}+x\frac{du}{dx}+\left(x^2-\nu ^2\right) u=0,
\label{7.14}
\end{equation}
but now with $\nu =1/5$. Eq. (\ref{7.14}) is the standard Bessel equation
\cite{AS72}, which have two independent solutions $J_\nu (x)$ and $Y_\nu (x)$. 
Therefore, the general solution of Eq.(\ref{7.6}) are given by
\begin{equation}
\psi =x^{1/2}\left(cJ_\nu (x)+dY_\nu (x)\right),  \label{7.15}
\end{equation}
where $c$ and $d$ are the integration constants, which will be determined
from the boundary conditions given by Eqs. (\ref{7.12a}) and (\ref{7.12b}).
Setting
\begin{eqnarray}
\Delta _{11} &\equiv &2J_\nu \left(x_c\right) -5x_cJ_{\nu +1}
\left(x_c\right),  \nonumber  \label{7.16} \\
\Delta _{12} &\equiv &2Y_\nu \left(x_c\right) -5x_cY_{\nu +1}
\left(x_c\right),  \nonumber \\
\Delta _{21} &\equiv &2J_\nu \left(x_0\right) -5x_0J_{\nu +1}
\left(x_0\right),  \nonumber \\
\Delta _{22} &\equiv &2Y_\nu \left(x_0\right) -5x_0Y_{\nu +1}
\left(x_0\right),
\end{eqnarray}
we find that Eqs. (\ref{7.12a}) and (\ref{7.12b}) can be cast in the form,
\begin{equation}
\left( \matrix{\Delta_{11} & \Delta_{12}\cr \Delta_{21} & \Delta_{22}\cr}%
\right) \left(\matrix{c \cr d\cr}\right) =0.  \label{7.17}
\end{equation}
It has non-trivial solutions only when
\begin{equation}
\Delta \equiv {\mbox{det}}\left( \Delta _{ij}\right) =0.  \label{7.18}
\end{equation}
Fig. \ref{roots} shows the solutions of $\Delta =0$ for $x_0=my_0=0.01,%
\;1.0,\;1000$, respectively. From this figure, two remarkable features are:
(1) The spectrum of the KK towers is discrete. (2) The KK modes weakly
depend on the specific values of $x_0$.

Table I shows the first three modes $m_{n}\; (n = 1, 2, 3)$ for $x_{0} =
0.01, \; 1.0, \; 1000$, from which we can see that to find $m_{n}$ it is
sufficient to consider only the case where $x_{0} \gg 1$.

\begin{table}[tbp]
\begin{tabular}{|c|c|c|c|}
\hline
\label{table1} $x_{0}$ & $m_{1}y_{c}$ & $m_{2}y_{c}$ & $m_{3}y_{c}$ \\ \hline
0.01 & 3.55 & 6.72 & 9.87 \\ \hline
1.0 & 3.25 & 6.41 & 9.56 \\ \hline
1000 & 3.14 & 6.28 & 9.42 \\ \hline
\end{tabular}
\caption{The first three modes $m_{n}\; (n = 1, 2, 3)$ for $x_{0} = 0.01, \;
1.0, \; 1000$, respectively.}
\end{table}

When $x_0\gg 1$ we find that $x_c=x_0+my_c\gg 1$, and that \cite{AS72}
\begin{eqnarray}
J_\nu (x) &\simeq &-Y_{\nu +1}(x)\simeq \sqrt{\frac 2{\pi x}}\cos \left(
x-\frac 7{20}\pi \right) ,  \nonumber  \label{7.19} \\
Y_\nu (x) &\simeq &J_{\nu +1}(x)\simeq \sqrt{\frac 2{\pi x}}\sin \left(
x-\frac 7{20}\pi \right) .
\end{eqnarray}
Inserting the above expressions into Eqs. (\ref{7.16}) and (\ref{7.18}), we
obtain
\begin{eqnarray}
\Delta  &=&-\sqrt{\frac 4{\pi ^2x_0x_c}}\left\{ 10\left(x_c-x_0\right) \cos
\left(x_c-x_0\right) \right.   \nonumber  \label{7.20} \\
&&\left. +\left( 4+25x_0x_c\right) \sin \left(x_c-x_0\right) \right\} ,
\end{eqnarray}
whose roots are given by
\begin{equation}
\tan \left(x_c-x_0\right) =-\frac{10\left(x_c-x_0\right) }{4+25x_0x_c}.
\label{7.21}
\end{equation}
From this equation, we can see that $m_n$ satisfies the bounds
\begin{equation}
\frac{n\pi }{y_c}<m_n<\frac{(n+1)\pi }{y_c},\;(n=1,2,3,...).  \label{7.22}
\end{equation}
Combining the above expression with Table I, we find that $m_n$ is well
approximated by
\begin{equation}
m_n\simeq n\pi \left( \frac{l_{pl}}{y_c}\right) M_{pl},  \label{7.23}
\end{equation}
For $x_0\gg 1$. In particular, we have
\begin{eqnarray}
m_1 &\simeq &3.14\times \left( \frac{10^{-19}\;{\mbox{m}}}{y_c}\right) \;{%
\mbox{TeV}}  \nonumber  \label{7.24} \\
&\simeq &\cases{1\; {\mbox{TeV}}, & $y_{c} \simeq 10^{-19} \;
{\mbox{m}}$,\cr 10^{-2} \; {\mbox{eV}}, & $y_{c} \simeq 10^{-5} \;
{\mbox{m}}$,\cr 10^{-4} \; {\mbox{eV}}, & $y_{c} \simeq 10^{-3} \;
{\mbox{m}}$.\cr}
\end{eqnarray}

It should be noted that the   mass $m_{n}$ calculated above is measured
by the observer with the metric $\eta_{\mu\nu}$. However, since the warped factor
$e^{\sigma(y)}$ is not one at $y = y_{c}$, the physical mass on the visible brane 
should be given by \cite{RS1}
\bq
\lb{phycialmass}
m^{obs}_{n} = e^{-\sigma\left(y_{c}\right)}m_{n}
= \left(\frac{y_{c} + y_{0}}{L}\right)^{1/5} m_{n}.
\eq
Without introducing any new hierarchy, we expect that 
$\left[({y_{c} + y_{0})}/{L}\right]^{1/5} \simeq {\cal{O}}(1)$. As a result, 
we have 
\bq
\lb{phycialmassb}
m^{obs}_{n}  
= \left(\frac{y_{c} + y_{0}}{L}\right)^{1/5} m_{n} \simeq m_{n}.
\eq

For each $m_{n}$ that satisfies Eq. (\ref{7.18}), the wavefunction $%
\psi_{n}(x)$ is given by
\begin{eqnarray}  \label{7.25}
\psi_{n}(x) &=& N_{n}x^{1/2}\left\{\Delta_{12}\left(m_{n}, y_{c}\right)
J_{\nu}(x) \right.  \nonumber \\
& & \left. - \Delta_{11}\left(m_{n}, y_{c}\right) Y_{\nu}(x)\right\},
\end{eqnarray}
where $N_{n} \equiv N_{n}\left(m_{n}, y_{c}\right)$ is the normalization
factor, so that
\begin{equation}  \label{7.26}
\int^{y_{c}}_{0}{\left|\psi_{n}(x)\right|^{2} dy} = 1.
\end{equation}

\subsection{4D Newtonian Potential and Yukawa Corrections}

To calculate the four-dimensional effective Newtonian potential and its
corrections, let us consider two point-like sources of masses $M_1$ and 
$M_2$, located on the brane at $y=y_c$. Then, the discrete eigenfunction $\psi
_n(x)$ of mass $m_n$ has an Yukawa correction to the four-dimensional
gravitational potential between the two particles \cite{BS99,Csaki00}
\begin{equation}
U(r)=G_4\frac{M_1M_2}r+\frac{M_1M_2}{M_5^3r}\sum_{n=1}^\infty {%
e^{-m_nr}\left| \psi _n(x_c)\right| ^2},  \label{7.27}
\end{equation}
where $\psi _n(x_c)$ is given by Eq. (\ref{7.25}). When $x_0=m_ny_0\gg 1$,
from Eq. (\ref{7.19}) we find that
\begin{eqnarray}
N_n &\simeq &\sqrt{\frac{\pi ^2}{50x_cy_c}},  \nonumber  \label{7.28} \\
\psi _n(x_c) &\simeq &\sqrt{\frac 2{y_c}}.
\end{eqnarray}
Then, it can be seen that all terms except for the first one in Eq. (\ref{7.27})
are exponentially suppressed, and have negligible contributions to the 4D
effective potential $U(r)$.


\section{Brane Cosmology}

\renewcommand{\theequation}{5.\arabic{equation}} \setcounter{equation}{0}

In this section, we shall apply the formulas developed in Section II
to cosmology.

\subsection{General Metric and Gauge Choices}

The general metric for cosmology takes the form \cite{WS07,WCS08},
\begin{equation}  
\label{4.1}
ds^{2}_{5} = g_{ab} dx^{a}dx^{b} =g_{MN} dx^{M}dx^{N} -
e^{2\omega\left(x^{M}\right)}d\Sigma^{2}_{k},
\end{equation}
where $M, \; N = 0, 1$, and
\begin{equation}  \label{4.1a}
d\Sigma^{2}_{k} = \frac{dr^{2}}{1 - k r^{2}} + r^{2}\left(d\theta^{2} +
\sin^{2}\theta d\varphi^{2}\right),
\end{equation}
where the constant $k$ represents the curvature of the 3-space, and can be
positive, negative or zero. Without loss of generality, we shall choose
coordinates such that $k = 0, \pm 1$. The metric (\ref{4.1}) is invariant
under the coordinate transformations,
\begin{equation}  \label{4.3}
{x^{\prime}}^{N} = f^{N}\left({x}^{M}\right).
\end{equation}
Using these two degrees of freedom, one can choose different gauges.

\subsubsection{The Canonical Gauge}

In particular, in \cite{WCS08} the gauge was chosen such that
\begin{equation}  \label{4.3b}
g_{01} = 0, \;\;\; y_{I} = 0,\; y_{c},
\end{equation}
where $y_{I}$ denote the locations of the two orbifold branes, with $y_{c}$
being a constant. Then, the general metric can be cast in the form,
\begin{equation}  \label{4.4a}
ds^{2}_{5} = N^{2}(t,|y|)dt^{2} - B^{2}(t,|y|)dy^{2} - B^{2}(t,|y|)
d\Sigma^{2}_{k},
\end{equation}
where $|y|$ is defined as that given in \cite{WCS08} [cf. Fig. \ref{fig1}].
By using distribution theory, the field equations on the two branes were
obtained explicitly in terms of the discontinuities of the metric
coefficients $A,\; B$ and $N$. For the details, we refer readers to \cite
{WCS08}. The gauge Eq. (\ref{4.3b}) will be referred to as \emph{the
canonical gauge}.

\subsubsection{The Conformal Gauge}

One can also choose the gauge
\begin{equation}  \label{4.3a}
g_{00} = g_{11}, \;\;\; g_{01} = 0,
\end{equation}
so that the general five-dimensional metric takes the form,
\begin{equation}  \label{4.4}
ds^{2}_{5} = e^{2\sigma(t,y)}\left(dt^{2} - dy^{2}\right) -
e^{2\omega(t,y)}d\Sigma^{2}_{k}.
\end{equation}
But with this gauge, the hypersurfaces of the two branes are not fixed, and
usually given by $y = y_{I}(t)$. We shall refer the gauge Eq. (\ref{4.3a}) to
as \emph{the conformal gauge}. It should be noted that in this conformal
gauge, metric (\ref{4.4}) still has the remaining gauge freedom,
\begin{equation}  
\label{4.5}
t = f\left(\xi_{+}\right) + g\left(\xi_{-}\right), \;\;\; y =
f\left(\xi_{+}\right) - g\left(\xi_{-}\right)
\end{equation}
where $\xi_{\pm} \equiv t^{\prime}\pm y^{\prime}$, and $f\left(\xi_{+}%
\right) $ and $g\left(\xi_{-}\right)$ are arbitrary functions of their
indicated arguments.

It should be noted that in \cite{Martin04} comoving branes were considered, 
and it was claimed that the gauge freedom of Eq. (\ref{4.5}) can always bring 
the two branes at rest (comoving). However, from Eq. (A5) of \cite{Martin04} 
it can be seen that this is not true (at least) at the moment of the collision, 
$\tilde{y}_{2}(t) = 0$, for which Eq. (A5) reduces to $f(t) = f(t) + 2$, 
which is not satisfied for any finite function $f(t)$. In addition, using 
Eq. (\ref{4.5}), one can always bring one brane at rest, as shown in \cite{Bowcock00} 
(See also  \cite{Martin04}). In this paper, we shall leave this possibility 
open,  and choose to work with the conformal gauge, in which the branes are
located on the surfaces $y = y_{I}(t)$.

\subsection{Field Equations Outside the Two Branes}

It can be shown that outside the two branes the field equations (\ref{3.1a})
for the metric (\ref{4.4}) have four independent components, which can be
cast in the form,
\begin{eqnarray}
\label{4.6a}
&&\omega _{,tt}+\omega _{,t}\left( \omega _{,t}-2\sigma _{,t}\right) +\omega
_{,yy}+\omega _{,y}\left( \omega _{,y}-2\sigma _{,y}\right)   \nb\\
&&\;\;\;\;\;\;\;\;\;\;\;\;\;\;\;=-\frac 16\left( {\phi _{,t}}^2+{\phi _{,y}}%
^2\right) ,  \\
\label{4.6b}
&&2\sigma _{,tt}+\omega _{,tt}-3{\omega _{,t}}^2-\left( 2\sigma
_{,yy}+\omega _{,yy}-3{\omega _{,y}}^2\right)  \nonumber \\
&&\;\;\;\;\;\;\;\;\;\;\;\;\;\;\;-4ke^{2(\sigma -\omega )}  \nonumber \\
&&\;\;\;\;\;\;\;\;\;\;\;\;\;\;\;=-\frac 12\left( {\phi _{,t}}^2-{\phi _{,y}}%
^2\right) , \\
\label{4.6c}
&&\omega _{,ty}+\omega _{,t}\omega _{,y}-\left( \sigma _{,t}\omega
_{,y}+\sigma _{,y}\omega _{,t}\right)  \nonumber \\
&&\;\;\;\;\;\;\;\;\;\;\;\;\;\;\;=-\frac 16\phi _{,t}\phi _{,y}, \\
\label{4.6d}
&&\omega _{,tt}+3{\omega _{,t}}^2-\left( \omega _{,yy}+3{\omega _{,y}}%
^2\right) +2ke^{2(\sigma -\omega )}  \nonumber \\
&&\;\;\;\;\;\;\;\;\;\;\;\;\;\;\;=2\alpha ^2e^{2(\sigma -\phi )}.
\end{eqnarray}

On the other hand, the Klein-Gordon equation (\ref{3.1b}) takes the form,
\bq
\label{4.8}
\phi _{,tt}+3\phi _{,t}\omega _{,t}-\left( \phi _{,yy}+3\phi _{,y}\omega
_{,y}\right)
 =12\alpha ^2e^{2(\sigma -\phi)}.
\eq

\subsection{Field Equations on the Two Branes}

Eqs. (\ref{4.6a}) - (\ref{4.8}) are the field equations that are valid in
between the two orbifold branes,
\bq
\lb{4.9a}
y_{2}(t) < y < y_{1}(t).
\eq
The proper distance between the two branes is given by
\begin{equation}  
\label{4.8c}
{\mathcal{D}}(t) = \int_{y_{2}}^{y_{1}}{e^{\sigma(t, y)}dy}.
\end{equation}

On each of the two branes, the metric reduces to
\begin{equation}  \label{4.9}
\left. ds^{2}_{5}\right|_{M^{(I)}_{4}} =
g^{(I)}_{\mu\nu}d\xi_{(I)}^{\mu}d\xi_{(I)}^{\nu} = d\tau_{I}^{2} -
a^{2}\left(\tau_{I}\right)d\Sigma^{2}_{k},
\end{equation}
where $\xi^{\mu}_{(I)} \equiv \left\{\tau_{I}, r, \theta, \varphi\right\}$,
and $\tau_{I}$ denotes the proper time of the I-th brane, defined by
\begin{eqnarray}  \label{4.10}
d\tau_{I} &=& e^{\sigma\left[t_{I}(\tau_{I}), y_{I}(\tau_{I})\right]} \sqrt{%
1 - \left(\frac{\dot{y}_{I}}{\dot{t}_{I}}\right)^{2}}\; dt_{I},  \nonumber \\
a\left(\tau_{I}\right) &\equiv& e^{\omega\left[t_{I}(\tau_{I}),
y_{I}(\tau_{I})\right]},
\end{eqnarray}
with $\dot{y}_{I} \equiv d{y}_{I}/d\tau_{I}$, etc. For the sake of
simplicity and without of causing any confusion, from now on we shall drop
all the indices ``I", unless some specific attention is needed. Then, the
normal vector $n_{a}$ and the tangential vectors $e^{a}_{(\mu)}$ are given,
respectively, by
\begin{eqnarray}  
\label{4.11}
n_{a} &=& 
e^{2\sigma}\left(- \dot{y}\delta^{t}_{a} + \dot{t}%
\delta^{y}_{a}\right),  \nonumber \\
n^{a} &=& -   \left(\dot{y}\delta^{a}_{t} + \dot{t}%
\delta^{a}_{y}\right),  \nonumber \\
e^{a}_{(\tau)} &=& \dot{t}\delta^{a}_{t} + \dot{y}\delta^{a}_{y}, \;\;\;
e^{a}_{(r)} = \delta^{a}_{r},  \nonumber \\
e^{a}_{(\theta)} &=& \delta^{a}_{\theta},\;\;\; e^{a}_{(\varphi)} =
\delta^{a}_{\varphi}.
\end{eqnarray}
Thus, we find that  
\begin{eqnarray}  
\label{4.12}
{\mathcal{G}}^{(5)}_{\mu\nu} &=& {\mathcal{G}}^{(5)}_{\tau}
\delta^{\tau}_{\mu} \delta^{\tau}_{\nu} - {\mathcal{G}}^{(5)}_{\theta}
\delta^{m}_{\mu} \delta^{n}_{\nu}g_{mn},  \nonumber \\
E^{(5)}_{\mu\nu} &=& E^{(5)}\left(3 \delta^{\tau}_{\mu} \delta^{\tau}_{\nu}
- \delta^{m}_{\mu} \delta^{n}_{\nu}g_{mn}\right),
\end{eqnarray}
where $m,\; n = r, \; \theta, \; \varphi$, and
\begin{eqnarray}  \label{4.13}
{\mathcal{G}}^{(5)}_{\tau} &\equiv& \frac{1}{3}e^{-2\sigma}\left({\phi_{,t}}%
^{2} - {\phi_{,y}}^{2}\right) - \frac{1}{24}\left(5\left(\nabla\phi%
\right)^{2} - 6 V_{5}\right),  \nonumber \\
{\mathcal{G}}^{(5)}_{\theta} &\equiv& \frac{1}{24}\left(8{\phi_{,n}}^{2} + 5
\left(\nabla\phi\right)^{2} - 6 V_{5}\right),  \nonumber \\
E^{(5)} &\equiv& \frac{1}{6}e^{-2\sigma}\left(\left(\sigma_{,tt} -
\omega_{,tt}\right) - \left(\sigma_{,yy} - \omega_{,yy}\right) \right.
\nonumber \\
& & \left. + k e^{2(\sigma - \omega)}\right),
\end{eqnarray}
with 
\begin{equation}  \label{4.13a}
V_{5} \equiv 6\alpha^{2}e^{-2\phi}.
\end{equation}
Then, it can be shown that the four-dimensional field equations on each of
the two branes take the form,
\begin{eqnarray}
 \label{4.14a}
H^{2} &+& \frac{k}{a^{2}} = \frac{8\pi G}{3}\left(\rho + \tau_{\phi}\right)
+ \frac{1}{3}\Lambda + \frac{1}{3}{\mathcal{G}}^{(5)}_{\tau} + E^{(5)}
\nonumber \\
& & + \frac{2\pi G}{3\rho_{\Lambda}}\left(\rho + \tau_{\phi}\right)^{2}, \\
 \label{4.14b}
\frac{\ddot{a}}{a} &=& - \frac{4\pi G}{3}\left(\rho +3p - 2
\tau_{\phi}\right) + \frac{1}{3}\Lambda  \nonumber \\
& & - E^{(5)} - \frac{1}{6}\left({\mathcal{G}}^{(5)}_{\tau} + 3{\mathcal{G}}%
^{(5)}_{\theta}\right) - \frac{2\pi G}{3\rho_{\Lambda}}\left[\rho\left(2\rho
+ 3p\right) \right.  \nonumber \\
& & \left. + \left(\rho + 3p - \tau_{\phi}\right) \tau_{\phi}\right],
\end{eqnarray}
where $H \equiv \dot{a}/{a}, , \; \Lambda \equiv \Lambda_{4},\; G \equiv
G_{4}$ and $\rho_{\Lambda} = \Lambda/(8\pi G)$.

On the other hand, from Eqs. (\ref{3.22}) and (\ref{2.3gca}) we find that
\bqn
 \label{4.14c}
& & \phi _{,n}^{(I)} = \epsilon_{I}
\left(\kappa^{2}_{5}\frac{\partial V^{(I)}_{4}}{\partial \phi}
- 6 \alpha \epsilon _Ie^{-\phi}
+ \frac{1}{2}\sigma^{(I)}_{\phi}\right),\;\;\\
\label{4.14d}
& & \left(\dot{\rho}^{(I)} + \dot{\tau}^{(I)}_{\phi}\right) + 3H^{(I)} 
\left(\rho^{(I)} + p^{(I)}\right)
= \Pi^{(I)}, \;\;  \;\;
\eqn
where $H^{(I)} \equiv \left[da\left(\tau_{I}\right)/d\tau_{I}\right]/
a\left(\tau_{I}\right)$,  and
\bq
\lb{4.14e1}
\Pi^{(I)} \equiv \frac{\epsilon _I}{\kappa^{2}_{5}} 
           \phi^{(I)}_{,\tau}\phi^{(I)}_{,n}.
\eq
From Eqs. (\ref{2.3ec}) and (\ref{4.11}), we also find that
\bq
\lb{4.14da}
\dot{\tau}^{(I)}_{\phi} = \frac{\phi^{(I)}_{,\tau}}{\kappa^{2}_{5}}
\left\{\kappa^{2}_{5}\frac{\partial V^{(I)}_{4}}{\partial \phi}
- 6 \alpha \epsilon _Ie^{-\phi}\right\}.
\eq
Then, Eqs. (\ref{4.14c}) and (\ref{4.14d}) can be written as
\bqn
 \label{4.14c1}
& & \dot{\tau}^{(I)}_{\phi} = \Pi^{(I)}  - Q^{(I)},\\
\label{4.14d1}
& & \dot{\rho}^{(I)}  + 3H^{(I)} \left(\rho^{(I)} + p^{(I)}\right) = Q^{(I)}, 
\eqn
where 
\bq
\lb{4.14e1a}
Q^{(I)} \equiv \frac{1}{2\kappa^{2}_{5}}\phi^{(I)}_{,\tau} \sigma^{(I)}_{\phi}.
\eq
When there is only  gravitational interaction between the scalar field and 
the perfect fluid, we have $\sigma _\phi ^{(I)} = 0$ [cf. Es.(\ref{2.3h})], 
and then the above equations reduce to
\bqn
 \label{4.14c2}
& & \dot{\tau}^{(I)}_{\phi} = \Pi^{(I)}, \; \left(Q^{(I)} = 0\right),\\
\label{4.14d2}
& & \dot{\rho}^{(I)}  + 3H^{(I)} \left(\rho^{(I)} + p^{(I)}\right)
= 0, \;  \left(Q^{(I)} = 0\right).  \; \;\; \;\;
\eqn

Eqs.(4.14a)-(4.14d) form the complete set of the field equations on the branes. 
However, in order to solve them, additional information from bulk is needed. 
In particular, the $E^{(5)}$ term represents the projection of the five-dimensional
Weyl tensor on the branes, while the ${\cal{G}}^{(5)}_{\tau}$ and ${\cal{G}}^{(5)}_{\theta}$
terms represent the contribution of the bulk scalar field.  When the bulk is
conformally flat,  $E^{(5)}$ vanishes.  The bulk scalar contributions
are in general always present, although in some particular cases, such contributions might be 
negligible. Then, from  Eqs.(4.14a) and (4.14b) we can see that the universe will
be asymptotically approaching to the $\Lambda$CDM model for $\rho, \; \tau_{\phi}
\ll 1$ at later time $a \gg 1$. It should be noted that in \cite{GWW07}, 
using the large extra dimensions, we showed that the effective cosmological 
constant $\Lambda$ can be lowered to its observational
value. In the early universe, the quadratic  terms of $\rho$ will dominate, 
and $H \propto \rho$, a  feature that is commonly shared by  brane-world models 
\cite{branes}.

\section{Conclusions and discussions}
\renewcommand{\theequation}{6.\arabic{equation}} \setcounter{equation}{0}

In this paper, we have systematically studied the   brane world in the 
in the framework of  the Horava-Witten heterotic M-Theory on $S^{1}/Z_{2}$
along the line set up by Lukas {\em et al}  \cite{HW96,LOSW99}. In particular,
after reviewing the model in Sec. II, and writing separately down the general 
gravitational and matter field equations both in the bulk and on the branes, 
In Sec. III, we have shown explicitly that   the radion is stable, by using the  
Goldberger-Wise mechanism \cite{GW99}. After working out the specific relation 
between the distance, $Y_{c}$, of the two branes and the radion $\varphi$, 
we have obtained an explicit form of the radion mass $m_{\varphi}$ in terms of 
the relevant parameters of the model [cf. Eq.(\ref{5.5m8})].  By properly 
choosing these parameters, it can be seen that the radion mass can be of the 
order of GeV. 

We have also shown that the gravity is localized on the visible (TeV) brane [cf. Sec. IV], 
in contrast to the RS1 model in which the gravity is localized on the Planck 
(hidden) brane \cite{RS1}. In addition, the spectrum of the gravitational KK
modes is discrete, and given explicitly by Eq.(\ref{7.23}), which can be in the order
of TeV. The corrections to the 4D Newtonian potential from the higher order 
gravitational KK modes are exponentially  suppressed and can be safely neglected
[cf. Eq.(\ref{7.27})]. 

To apply  such a setup to cosmology, we have first found the general form
of metric, by embedding a constant curvature 3-space into a five-dimensional bulk,
and discussed the gauge conditions in details. Working with the conformal gauge,
we have written down the general gravitational and matter field equations in the bulk, 
given by Eqs.(\ref{4.6a})-(\ref{4.8}), and the generalized Friedmann-like 
equations on each of the two orbifold  branes, given by Eqs.(\ref{4.14a}) and 
(\ref{4.14b}). The conservation laws for the scalar and matter fields are given, 
respectively, by  Eqs.(\ref{4.14c}) and (\ref{4.14d}). These consist of the
complete set of field equations of the brane cosmology in the framework of  
the Horava-Witten heterotic M-Theory on $S^{1}/Z_{2}$ along the line set up by Lukas 
{\em et al}  \cite{HW96,LOSW99}.

In the study of brane worlds, one   of the most attractive features is that it may 
 resolve the long standing hierarchy problem, namely the large difference 
in magnitudes between the Planck and electroweak scales,
${M_{pl}}/{M_{EW}} \simeq 10^{16}$, 
where 
$M_{EW}$  denotes the electroweak scale with $M_{EW} \sim  TeV$.
In particular, using the large extra dimensions, ADD found that in a D-dimensional
bulk with the Planck scale $M_{D}$, the deduced 4-dimensional Planck mass $M_{pl}$
\cite{ADD98} is given by 
$M_{pl}^{2} = V_{D-4}M_{D}^{D-2}$, 
where $V_{D-4}$ denotes the volume of the extra (D-4)-dimensional space. Clearly, if
the extra dimensions are large enough, even $M_{D}$ is in the order of electroweak 
scale $M_{D} \simeq M_{EW} \simeq TeV$, one can get the correct order of $M_{pl} \simeq 
10^{16}\;TeV$, whereby the hierarchy problem is resolved. 
In the RS1 model, the mechanism is completely different \cite{RS1}. Instead of using large
dimensions, RS used the warped factor $\sigma(y) = k |y|$, for which the mass $m_{0}$\
measured on the invisible (Planck) brane is related to the mass $m$ measured on the visible 
(TeV) brane by $m = e^{-ky_{c}}m_{0}$. Clearly, by properly choosing the distance $y_{c}$ 
between the two branes, one can lower $m$ to the order of $TeV$, even $m_{0}$ is still in
the roder of $M_{pl}$. It should be noted that the five-dimensional Planck mass $M_{5}$
in the RS1 sceanrio is still of the order of $M_{pl}$ and the two are related by
$ M^{2}_{pl} = {M^{3}}{k^{-1}}\left(1 - e^{-2ky_{c}}\right) \simeq M^{2}_{5}$ for
$k \simeq M_{5}$.

It is important to note that, when deriving the relation between  $M_{D}$ and  $M_{pl}$, 
in both scenarios it was implicitly assumed that the 4-dimensional effective Einstein-Hilbert 
action $S_{g}^{eff.}$  couples with matter directly as
\bq
\lb{6.3}
S_{g}^{eff.} + S_{m} = 
\int{\sqrt{-g}d^{4}\left(-\frac{1}{2\kappa^{2}_{4}} R + {\cal{L}}_{m}\right)},
\eq
from which one obtains the Einstein field equations, $G_{\mu\nu} = \kappa^{2}_{4}\tau_{\mu\nu}$.
In the weak field limit, one arrives at $\kappa^{2}_{4} = 8\pi G/c^{4}$ \cite{Inverno}. However, in the 
brane-world scenarios, the coupling between the curvature and matter is much more complicated 
than that given by the above equation. In particular,  the gravitational field equations 
on the branes are given by Eq.(\ref{3.9}), which is a second-order polynomial in terms of the 
energy-momentum tensor $\tau_{\mu\nu}$ of the brane. In the weak-field regime, the quadratic
terms are negligible, and the term linear to  $\tau_{\mu\nu}$ dominates. Then, under the 
weak-field limit, one can show that $\kappa^{2}_{4}$ defined by Eq.(\ref{3.11}) is related
to the Newtonian constant exactly by  $\kappa^{2}_{4} = 8\pi G/c^{4}$, from which we find
that 
\bq
\lb{6.4}
g_{k} = \frac{6\kappa^{2}_{4}}{\kappa^{4}_{5}}.
\eq
Note that this result is quite general, and applicable to a large class of brane-world scenarios 
\cite{branes}. In the present case, we have $\kappa^{2}_{5} = M^{-3}_{5} = 1/(M^{9}_{11}
R^{6})$, where $R$ is the typical size of the extra dimensions \cite{GWW07}. Then, one find 
that $g_{k}  \simeq 10^{-47}\; GeV^{4}$, that is, to solve the hierarchy problem in the framework
of the HW heterotic M Theory on $S^{1}/Z_{2}$, the tension of the brane has to be in the 
same order of the current matter $\rho_{m}$, as well as of the observational cosmological 
constant $\rho_{\Lambda}^{obs}$, of course.

Finally, we would like to note that, when we considered the radion stability,  the 
backreaction of the Goldberger-Wise field $\Phi$ was not taken into account. 
In the Randall-Sandrum model \cite{RS1}, it was shown that such 
effects do not change the main conclusions of the stability of radion \cite{DeWolfe}.
It would be very interesting to show that it is also the case here.
It is also very important to   study  constraints  from other physical 
considerations, such as  the solar system tests, the formation of large-scale
structure, and the early universe.

\begin{acknowledgments}
YG thanks the hospitality of Baylor University where this work was
completed and the support by NNSFC under Grant No. 10605042. AW $\&$ QW 
thank the suport by NNSFC under Grant,  No. 10703005 and No. 10775119.
 
\end{acknowledgments}

\end{document}